\documentclass{aa}
\usepackage{natbib}
\bibpunct{(}{)}{;}{a}{,}{,}
\usepackage{amsmath}
\usepackage{graphicx}
\usepackage{txfonts}
\usepackage{color}
\usepackage[super]{nth}
\usepackage{ulem}

\begin{document}
\title{The VIMOS Public Extragalactic Redshift Survey}
\subtitle{Searching for cosmic voids \thanks{The voids catalogue (Table \ref{voidtable}) is only
    available at the CDS via anonymous ftp to cdsarc.u-strasbg.fr
    (130.79.128.5) or via
    http://cdsarc.u-strasbg.fr/viz-bin/qcat?J/A+A/XXX/AYY}}

\author{D.~Micheletti\inst{1}
\and A.~Iovino\inst{1}
\and A.~J.~Hawken\inst{1} 
\and B.~R.~Granett\inst{1}
\and M.~Bolzonella\inst{2}           
\and A.~Cappi\inst{2,21}
\and L.~Guzzo\inst{1}
\and U.~Abbas\inst{3}
\and C.~Adami\inst{4}
\and S.~Arnouts\inst{4}
\and J.~Bel\inst{1}
\and D.~Bottini\inst{5}
\and E.~Branchini\inst{6, 22,23}
\and J.~Coupon\inst{7}
\and O.~Cucciati\inst{2}           
\and I.~Davidzon\inst{2,11}
\and G.~De Lucia\inst{8}
\and S.~de la Torre\inst{4}
\and A.~Fritz\inst{5}
\and P.~Franzetti\inst{5}
\and M.~Fumana\inst{5}
\and B.~Garilli\inst{5}     
\and O.~Ilbert\inst{4}
\and J.~Krywult\inst{9}
\and V.~Le Brun\inst{4}
\and O.~Le F\`evre\inst{4}
\and D.~Maccagni\inst{5}
\and K.~Ma{\l}ek\inst{10,5}
\and F.~Marulli\inst{11,12,2}
\and H.~J.~McCracken\inst{13}
\and M.~Polletta\inst{5}
\and A.~Pollo\inst{14,15}
\and C.~Schimd\inst{4}
\and M.~Scodeggio\inst{5} 
\and L.~A.~M.~Tasca\inst{4}
\and R.~Tojeiro\inst{16}
\and D.~Vergani\inst{17,2}
\and A.~Zanichelli\inst{18}
\and A.~Burden\inst{16}
\and C.~Di Porto\inst{2}
\and A.~Marchetti\inst{5} 
\and C.~Marinoni\inst{19}
\and Y.~Mellier\inst{13}
\and T.~Moutard\inst{4}
\and L.~Moscardini\inst{11,12,2}
\and R.~C.~Nichol\inst{16}
\and J.~A.~Peacock\inst{20}
\and W.~J.~Percival\inst{16}
\and G.~Zamorani\inst{2}
}
\offprints{Daniele Micheletti \\ \email{daniele.micheletti@brera.inaf.it}}
\institute{INAF - Osservatorio Astronomico di Brera, Via Brera 28, 20122 Milano, via E. Bianchi 46, 23807 Merate, Italy 
\and INAF - Osservatorio Astronomico di Bologna, via Ranzani 1, I-40127, Bologna, Italy 
\and INAF - Osservatorio Astronomico di Torino, 10025 Pino Torinese, Italy 
\and Aix Marseille Universit\'e, CNRS, LAM (Laboratoire d'Astrophysique de Marseille) UMR 7326, 13388, Marseille, France  
\and INAF - Istituto di Astrofisica Spaziale e Fisica Cosmica Milano, via Bassini 15, 20133 Milano, Italy
\and Dipartimento di Matematica e Fisica, Universit\`{a} degli Studi Roma Tre, via della Vasca Navale 84, 00146 Roma, Italy 
\and Institute of Astronomy and Astrophysics, Academia Sinica, P.O. Box 23-141, Taipei 10617, Taiwan
\and INAF - Osservatorio Astronomico di Trieste, via G. B. Tiepolo 11, 34143 Trieste, Italy 
\and Institute of Physics, Jan Kochanowski University, ul. Swietokrzyska 15, 25-406 Kielce, Poland 
\and Department of Particle and Astrophysical Science, Nagoya University, Furo-cho, Chikusa-ku, 464-8602 Nagoya, Japan 
\and Dipartimento di Fisica e Astronomia - Universit\`{a} di Bologna, viale Berti Pichat 6/2, I-40127 Bologna, Italy 
\and INFN, Sezione di Bologna, viale Berti Pichat 6/2, I-40127 Bologna, Italy 
\and Institute d'Astrophysique de Paris, UMR7095 CNRS, Universit\'{e} Pierre et Marie Curie, 98 bis Boulevard Arago, 75014 Paris, France 
\and Astronomical Observatory of the Jagiellonian University, Orla 171, 30-001 Cracow, Poland 
\and National Centre for Nuclear Research, ul. Ho\.{z}a 69, 00-681 Warszawa, Poland 
\and Institute of Cosmology and Gravitation, Dennis Sciama Building, University of Portsmouth, Burnaby Road, Portsmouth, PO1 3FX 
\and INAF - Istituto di Astrofisica Spaziale e Fisica Cosmica Bologna, via Gobetti 101, I-40129 Bologna, Italy 
\and INAF - Istituto di Radioastronomia, via Gobetti 101, I-40129, Bologna, Italy 
\and Centre de Physique Th\'eorique, UMR 6207 CNRS-Universit\'e de Provence, Case 907, F-13288 Marseille, France 
\and SUPA, Institute for Astronomy, University of Edinburgh, Royal Observatory, Blackford Hill, Edinburgh EH9 3HJ, UK 
\and Laboratoire Lagrange, UMR7293, Université de Nice SophiaAntipolis, CNRS, Observatoire de la Cote d’Azur, 06300 Nice, France
\and INFN, Sezione di Roma Tre, via della Vasca Navale 84, I-00146 Roma, Italy 
\and INAF - Osservatorio Astronomico di Roma, via Frascati 33, I-00040 Monte Porzio Catone (RM), Italy 
}
\date{Received 30 April 2014; accepted 11 August 2014}

\abstract {The characterisation of cosmic voids gives unique
information about the large-scale distribution of galaxies, their
evolution, and the cosmological model.}
{We identify and characterise cosmic voids in the VIMOS Public
Extragalactic Redshift Survey (VIPERS) at redshift $0.55 < z <
0.9$.}
{A new void search method is developed based upon the identification
of empty spheres that fit between galaxies. The method can be used
to characterise the cosmic voids despite the presence of complex
survey boundaries and internal gaps. We investigate the impact of
systematic observational effects and validate the method against
mock catalogues. We measure the void size distribution and the
void-galaxy correlation function.}
{We construct a catalogue of voids in VIPERS. The distribution of
voids is found to agree well with the distribution of voids found in
mock catalogues. The void-galaxy correlation function shows
indications of outflow velocity from the voids.}
{}
{}
\maketitle

\section{Introduction}

Galaxies are distributed in a structure formed by filaments, walls,
groups and clusters that surround large regions of very low density:
the cosmic voids. This lack of any uniform population of `field
galaxies' gradually came into focus via the pioneering surveys of the
1970s \citep{chincarinirood76, tifftgregory76, tarenghietal78,
  gregorythompson78, chincarini78, joeveeretal78, tullyfisher78,
  chincarinirood79}. 

\citet{gregorythompson78} were the first to use the specific term
`void' to treat low-density regions as specific objects; see
\citet{Rood1988a} and \citet{Rood1988b} for a comprehensive review and
historical account of those years. The phenomenology of voids became
more clearly defined with the more extensive projects of the early
1980s, most notably the CfA I \citep{Davisetal1982} and CfA II surveys
\citep{deLapparentetal1986}, when these regions were described as
components of the structure of the universe.

There has been a resurgence of interest in cosmic voids over recent
years. This has been the result of high quality photometric and
spectroscopic data, combined with a strong theoretical motivation.  In
particular, void searches in the Sloan Digital Sky Survey (SDSS) Data
Release 7 main galaxy and luminous red galaxies (LRG) samples have
proved fruitful \citep{sutteretal12}, providing a rich resource for
numerous studies. These voids have been found to have an intricate
sub-structure \citep{alspaslanetal14}, to exhibit a significant
gravitational lensing signal \citep{clampittjain14}, and to produce
measurable cold spots on the CMB \citep{ilicetal13}.

Properties of voids, such as their internal structure, shape, and
distribution, are strongly dependent on the cosmological
scenario. Therefore, the study of voids can provide useful information
that can be used to put constraints on global cosmological parameters
\citep{ryden95,betancourtetal09} and evaluate proposed cosmological
models
\citep{icke84,regosgeller91,bensonetal03,colbergetal05,lavauxwandelt10,biswasetal10,hernandezsmith12,ceccarellietal13}.

As the most under-dense regions of the universe, voids are vacuum
dominated regions. Subsequently, they are excellent places within
which to test theories of dark energy
\citep{parklee07,bosetal12}. Some theories of modified gravity propose
that gravitation should work differently over large scales and in very
low density environments. Because voids are not only extremely
under-dense, but are also the largest components of structure in terms
of comoving size, they are well suited to studies of modified gravity
\citep{martinsheth09, sployaretal13, clampittetal13}. Important
information can also be obtained from redshift-space distortions
connected to galaxy outflow from the inside of the voids towards
over-dense regions surrounding them \citep{dekelrees94,rydenmelott96}.

There are different approaches to defining voids in the
literature. For example, volumes, from which galaxies brighter or more
massive than a given threshold are absent, can be used to estimate the
Void Probability Function \citep{1979MNRAS.186..145W}. In other cases,
such as ours, voids are defined as volumes that are not completely
empty but where the density contrast is below some threshold.

In this paper, we describe a void-finding algorithm and its
application to data from VIPERS (VIMOS Public Extragalactic Redshift
Survey; \citealp{guzzoetal13}). VIPERS offers a unique opportunity to
look for cosmic voids at higher redshift thanks to its wide angular
coverage and efficient sampling rate.

\begin{figure*}[!ht] 
\centering
\includegraphics[width=\textwidth]{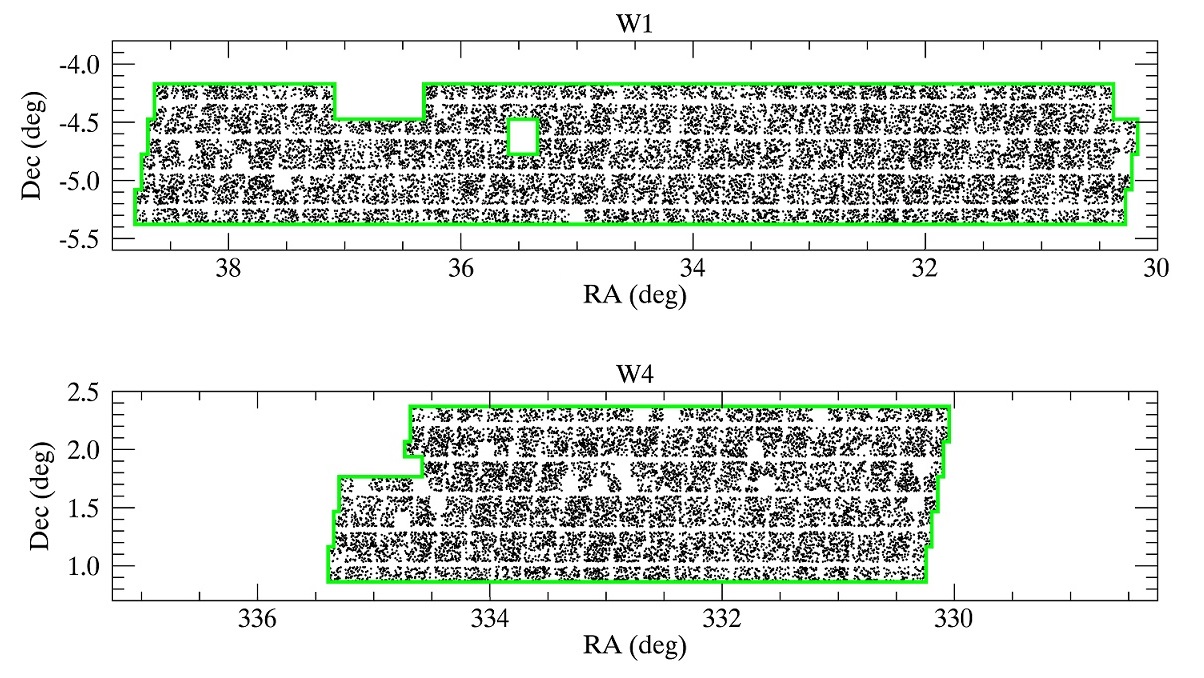}
\caption{RA-Dec distribution of galaxies in the W1 and W4 VIPERS
  fields for the internal catalogue version 4.0 used in this study
  (see text). The green lines are the survey borders that we consider
  in this paper.}
\label{w1w4borders}
\end{figure*} 

We structure the paper as follows:
in Sec. 2 we briefly describe the VIPERS survey and the samples we derived from the data;
in Sec. 3, we describe the mock galaxy catalogues used in this analysis;
in Sec. 4, we describe the tests carried out to evaluate the impact of observational effects on low-density regions detection process;
in Sec. 5, we describe the void-search algorithm and its application to VIPERS;
in Sec. 6, we present the void-galaxy correlation function; and 
Sec. 7, gives our discussion of the results and conclusions.

In this paper, we adopt standard cosmological parameters matching the
MultiDark simulations \citep{pradaeetal12} with $\Omega_m$ = 0.27,
$\Omega_{\Lambda}$ = 0.73 and $H_0$ = 70 $km \cdot s^{-1} \cdot
Mpc^{-1}$.

\section{The VIMOS Public Extragalactic Redshift Survey}
The VIMOS Public Extragalactic Redshift Survey (VIPERS) is an ongoing
ESO programme with the goal of studying the large scale distribution of
galaxies at moderate redshift (z = 0.5 - 1.2)
\citep{guzzoetal13,garillietal14}. VIPERS is constraining the
cosmological growth of structure as well as the properties of galaxies
and their environments. Results include measurements of the rate of
structure formation through redshift-space galaxy clustering at
$z=0.8$ \citep{delatorreetal13}, the dependence of galaxy clustering
on luminosity and stellar mass \citep{marullietal13} and the galaxy
stellar mass and luminosity functions
\citep{davidzonetal13,fritzetal14}.  The survey targets galaxies
within two regions of the W1 and W4 fields of the CFHTLS-Wide Survey
(Canada-France-Hawaii Telescope Legacy Wide;
\citealp{cuillandreetal12}), shown in Fig. \ref{w1w4borders}. By the
end of the survey, VIPERS will cover a total sky area of $24 \, {\rm
  deg}^2\!$, so that the survey area, redshift range, and relatively high
sampling rate translate into similar volumes and number densities as
local surveys such as 2dFGRS \citep{collessetal03}, SDSS
\citep{straussetal02} and GAMA \citep{driveretal11} at $z=0.3$.

We carried out the observations using the VIMOS (VIsible
Multi-Object Spectrograph) instrument at ESO-VLT (European Southern
Observatory - Very Large Telescope; \citealp{lefevretal03}). VIMOS is
a four-channel spectrograph, with each channel corresponding to a
quadrant of 7 by 8 ${\rm arcmin}$. The position of the four
quadrants leaves an unobserved cross-shaped area (of $\sim2$ arcmin
width) which is seen imprinted in the distribution of observed
galaxies (see Fig. \ref{w1w4borders}).

Galaxies have been targeted from the CFHTLS-Wide optical photometric
catalogue with an apparent magnitude limit set to $i_{AB} \leq
22.5$. A colour selection has been used to exclude galaxies at
redshift $z < 0.5$ \citep{guzzoetal13}.  Only a fraction of the
galaxies selected as potential targets may be observed because of
instrumental constraints, and of these, not all are guaranteed to give
a reliable redshift measurement.  Nevertheless, the combination of
target selection and observing strategy provides an overall sampling
rate of 35\%, which is extremely valuable for studying the
distribution of galaxies on moderate scales.

For each galaxy, the $B$-band rest-frame magnitude was estimated
following the Spectral Energy Distribution (SED) fitting method
described in \citet{davidzonetal13} and adopted to define volume
  limited samples. The choice of $B$-band rest-frame is natural,
  corresponding to the observed $I$-band at redshift $\sim 0.8$.  We derived
$k$ and colour corrections from the best-fitting SED
templates using all available photometry including near-UV, optical, 
and near-infrared.

\begin{figure}[!ht] 
\centering
\includegraphics[width=9cm]{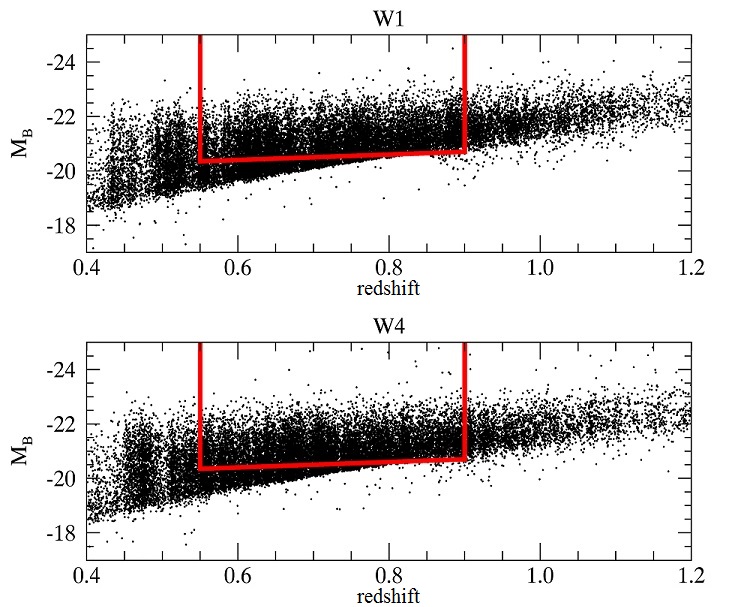}
\caption{Samples obtained form VIPERS data for the W1 and W4 fields.}
\label{Vsmpl}
\end{figure} 

\subsection{Galaxy samples}
The first public data release of VIPERS was made available in October
2013 and included 64\% of the final dataset \citep{garillietal14}.  In
this study we use the internal release version 4.0 catalogue, which
increases the area of the W1 field by 30\% and represents 75\% of the
final coverage.  The angular size of the fields are $7.4 \, {\rm
  deg}^2$ in W4 and $9.9 \, {\rm deg}^2$ in W1.  We select galaxies
with confident spectroscopic redshifts, corresponding to {\tt zflag}
$\ge 1.5$, yielding 34965 objects in W1 and 27258 in W4. For a
detailed description of the meaning of VIPERS flag values see
\citet{guzzoetal13}.

We wish to perform a consistent void search across the redshift range
of the survey. Therefore, we extracted two B-band rest-frame
volume-limited samples from the VIPERS catalogue, one for each field
(W1 and W4). As a result of galaxy evolution, the global galaxy
luminosity function varies with time \citep{ilbertetal05,zuccaetal09}
and the magnitude cut must therefore account for this evolution, as
shown in figure \ref{Vsmpl}. The chosen redshift range is $0.55 <z<
0.9$ giving a compromise between volume explored and observed galaxy
number density. The adopted magnitude limit is defined in the B-band
rest-frame as
\begin{equation}
M_{B,lim}(z) = M_{B,lim}(z = 0) - z ,
\end{equation}
where $M_{B,lim}(z)$ is the magnitude limit at redshift $z$. In the
present case, the chosen value for $M_{B,lim}$ is $-19.8$ at $z = 0$
(from now on, $M_{B,lim}$ refers to the value of the magnitude cut at
$z = 0$).

By definition the galaxy density is constant within a volume limited
sample. However, in reality, this may not be exactly the case because
of observational biases. Therefore we measure the mean inter-galaxy
separation, within the survey mask, as a function of redshift. The
mean inter-galaxy separation is defined as:
\begin{equation} \label{eq:medsep}
d_{sep} = 2\left(\frac{V}{\frac{4}{3} \pi n_g}\right)^{\frac{1}{3}};
\end{equation}
where $V$ is the survey volume, $n_g$ is the number of galaxies, and
$d_{sep}/2$ is the radius of the volume associated with each galaxy.

\begin{figure}[!ht] 
\centering
\includegraphics[width=9cm]{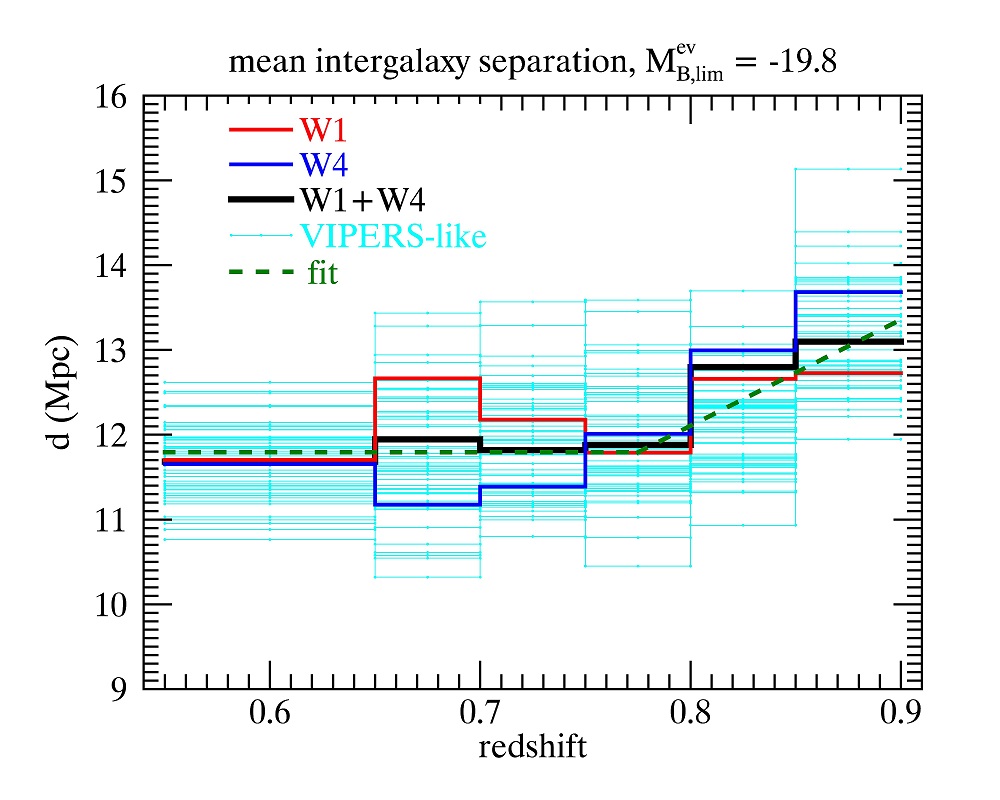}
\caption{Mean inter-galaxy separation for the W1 and W4 volume-limited
  samples (red and blue lines, combined value in black) and for mock
  survey mask samples (light blue lines, individual values). The green
  dashed line is the fit for the combined mean inter-galaxy separation
  of W1 and W4 fields.}
\label{dsep1}
\end{figure}

We measure $d_{sep}$ for the two fields in six redshift bins with
edges at $z = [0.55,0.65,0.70,0.75,0.80,0.85,0.90]$ (see
Fig. \ref{dsep1}). We observe that the mean separation is not constant
but increases significantly in the two highest redshift bins
($z=0.8-0.85$ and $z=0.85-0.9$). The trend can be explained by the
survey completeness that was found to decrease at fainter apparent
magnitude, $i_{AB} > 20.5$, due to the degradation of the redshift
measurement success rate \citep{guzzoetal13}. This incompleteness of
galaxies with the faintest apparent magnitudes predominantly affects
the highest redshift bins in a volume-limited sample. We account for
this effect when constructing the mock catalogues, as discussed in
Sec. 3.

\section{Mock catalogues} \label{MC}

To test the effects of observational biases on the detection of voids, 
we have used mock galaxy catalogues created to have the same
characteristics as the VIPERS data.

The mock catalogues are obtained from the MultiDark dark matter N-body
simulation, with cosmological parameters $\Omega_m = 0.27$, $\Omega_b
= 0.0469$, $\Omega_{\Lambda} = 0.73$, $H_0 = 70 km \cdot s^{-1} \cdot
Mpc^{-1}$, $n_s = 0.95$, and $\sigma_8 = 0.82$
\citep{pradaeetal12}. The dark matter halos have been populated with
galaxies using the Halo Occupation Distribution calibrated with VIPERS
data. The construction of the mock catalogues can be found in
\citet{delatorreetal13}. The resulting mock galaxy catalogues include
26 independent realisations of the W1 survey field and 31 realisations
of the W4 field.  These catalogues have a redshift range of $0.4 < z <
1.3$ and the same apparent magnitude limit as VIPERS ($i_{AB} <
22.5$). We replicated the VIPERS target selection procedure on the
mock catalogues obtaining the following samples:
\begin{itemize}
\item \textit{100\% mocks}: containing all mock galaxies brighter than
  the chosen absolute magnitude limit within the survey border
  (indicated by the green lines in Fig. \ref{w1w4borders}) but without
  internal gaps.
\item \textit{100\% mocks real space}: the same as above but using
  the real space positions of galaxies.
\item \textit{VIPERS-like mocks}: including the survey mask and
  accurately reproducing the galaxy number density in the VIPERS
  volume limited sample that we considered.
\end{itemize}

We matched the galaxy number density in the mocks to the VIPERS data
taking into account the apparent magnitude-dependent sampling rate
described in \citet{guzzoetal13}. The mocks well reproduce the trends
observed in the real data. This is shown in Fig. \ref{dsep1} where
the mean inter-galaxy separation measurements in the VIPERS-like mocks
have been over-plotted with thin blue lines.  The adopted apparent
magnitude dependent sampling rates are:

\begin{equation*}
SR(m) = \begin{cases} $95 \%$& \text{for } 21 < i_{AB} \leq 21.5 \\ $89 \%$& \text{for } 21.5 < i_{AB} \leq 22.0 \\ $84 \%$& \text{for } 22.0 < i_{AB} \leq 22.5 . \end{cases}
\end{equation*}

\section{Tests on mock catalogues} \label{MT}
In order to ensure that the search for voids is unhampered by the
peculiar survey geometry of VIPERS, masking effects such as the gaps
in each VIMOS pointing, and redshift space distortions, extensive
tests were performed on the mock catalogues described in the previous
section.

Galaxy counts-in-cells are a popular technique for analysing density
variations in surveys \citep{1980lssu.book.....P} and they also
provide a flexible tool for investigating the impact of complex
selection effects.  Furthermore it is common to use counts-in-cells to
search for scarcely populated regions that are associated with voids
(e.g. \citealt{hoylevogeley02}).  By comparing the galaxy
counts-in-cells of mocks without systematic effects to those with
systematic effects added (VIPERS-like), we can evaluate the impact of
observational biases (from the mask and missing quadrants) and
redshift space distortions on the observed galaxy spatial
distribution. We also refer the reader to \citet{cucciatietal14} for a
detailed study of the impact of observational systematics on the
density field estimate.

We chose the cell size to match the sizes of voids that may be found
in VIPERS. As will be discussed in Sec. \ref{ADstatsig}, the void
sizes that can be reliably probed in the VIPERS data range in radius
from $15 \lesssim r \lesssim 30$Mpc.

To compute the counts-in-cells, we randomly place a sufficient number
of spherical cells inside the unmasked and undistorted mock to
massively over-sample the available volume, keeping only those for
which at least 80\% of the volume is inside the survey borders (green
lines in Fig. \ref{w1w4borders}). The amount of overlap allowed
between the cells and the survey boundaries is somewhat arbitrary, we
find that a 20\% limit provides us with enough cells to calculate
robust statistics whilst not being so conservative as to dramatically
reduce the effective survey volume.  We then place cells at the same
position in the second (masked or distorted) mock so that we can
compare counts at the same position in space in the two cases.

We then identify the cells at the extremities of the histogram of the
probability density function (PDF), below the 15th and above the 85th
percentile for one of the samples, and then we check where these cells
are placed in the distribution of the other sample. This is done for
both distributions, checking where the extreme cells of the first
sample are placed in the distribution of the second sample and where
the extreme cells of the second sample are placed in the distribution
of the first sample.

\subsection{Real and redshift space}
The redshift measured in a galaxy survey may be considered as a
recessional velocity and is used as a proxy for distance. The measured
recessional velocity is, however, the sum of a component caused by the
universe's expansion (cosmological redshift) and a proper motion
component caused by gravitational interaction. This alteration of the
observed positions of galaxies can thus affect both the shape and size
of observed voids.

There are two types of redshift space distortion that can affect the
detection of low-density regions.  Firstly, linear redshift space
distortions (outflow): galaxies in low-density regions are subjected
to gravitational attraction from higher-density structures, and their
observed position is altered such that they appear nearer to the
structure than they really are \citep{kaiser86}. This has the effect
of further emptying the low-density regions.  Secondly, non-linear
redshift space distortions (the so-called fingers-of-god effect):
galaxies in relaxed high-density regions (clusters) have a velocity
dispersion determined by the gravitational potential well of the
structure. This has the effect of stretching over densities in
redshift space along the line of sight and possibly moving the
observed position of some galaxies into lower-density regions
\citep{1972MNRAS.156P...1J}.

To evaluate how redshift space distortions affect the observed
distribution of galaxies we performed counts-in-cells tests on the
100\% mock sample, in redshift space and real space.  Examining the
rank of cells in the PDF when selected in redshift space and viewed in
real space (and vice versa) allows us to check if regions that we
observe as being at high or low density in redshift space are truly so
in real space.

\begin{figure}[t] 
\centering
\includegraphics[width=0.5\textwidth]{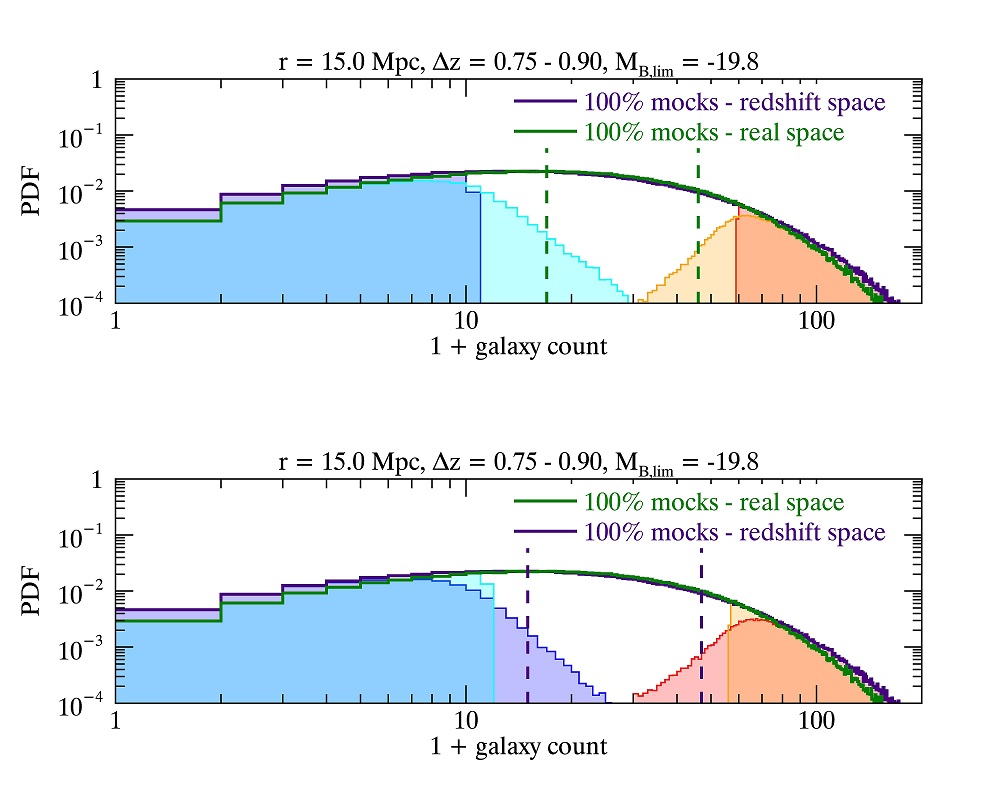}
\caption{Counts-in-cells distributions in the mock samples in redshift
  bin $0.75 < z <0.9$ for cells of radius $r = 15$ Mpc. The green line
  shows the distribution of counts in real space while the violet line
  shows the distribution in redshift space (extending slightly to
  higher counts). Top panel: cells are selected in redshift
  space. Blue and red shaded regions represent the 15\% least and most
  populated cells in the redshift space sample, respectively. In real
  space the distribution of these cells are shown by the cyan and
  orange shaded regions.  The vertical dashed line indicates the 25th
  and 75th percentiles of the real space distribution.  Bottom panel:
  vice versa, cells are selected in real space and viewed in redshift
  space.  The vertical dashed line indicates the 25th and 75th
  percentiles of the redshift space distribution.}
\label{ZRccount}
\end{figure} 

We expect redshift space distortions to have a bigger impact on
smaller voids. Therefore, we show histograms for counts-in-cells for a
cell radius of $r = 15$ Mpc, Fig. \ref{ZRccount}, in the redshift
interval $0.75 < z < 0.9$. The histograms show that more than 95\% of
cells that contribute to the 15th percentile tails of the redshift
space distribution remain in the tails after transforming to real
space, and, similarly, the tails selected in real space remain in the
tails in redshift space.  Thus, the extremities of the distribution are
preserved and we can be confident that the true low-density regions in
real space can be detected in redshift space.

\subsection{Observational biases}
\begin{figure}[h] 
\centering
\includegraphics[width=9cm]{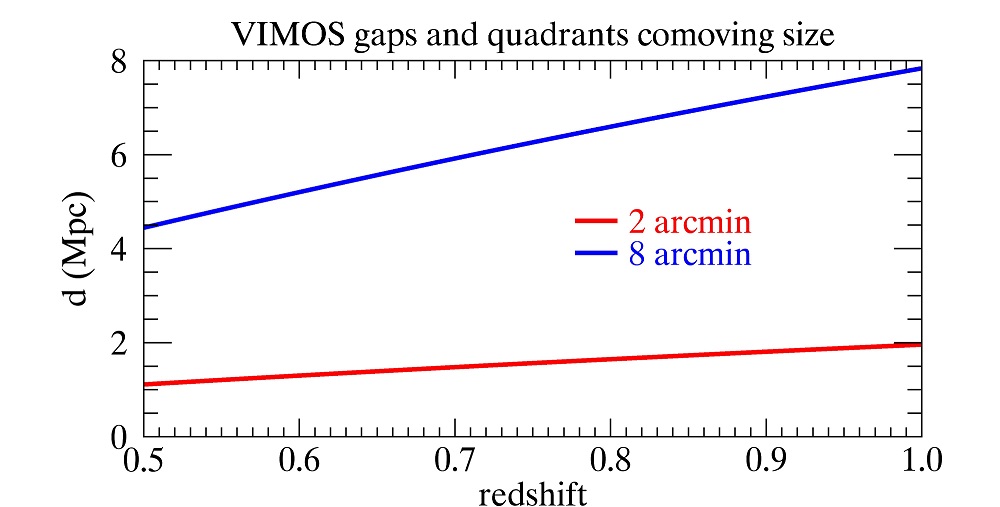}
\caption{Transverse comoving sizes of VIMOS mask cross-shaped gaps (2
  arcmin) and quadrants (8 arcmin) with redshift. The sizes, in the
  redshift range of interest, are smaller than the dimensions of the
  sphere used for counts-in-cells, so we do not expect a large impact
  on the counts.}
\label{VIMOSgaps}
\end{figure} 

\begin{figure}[!h] 
\centering
\includegraphics[width=0.5\textwidth]{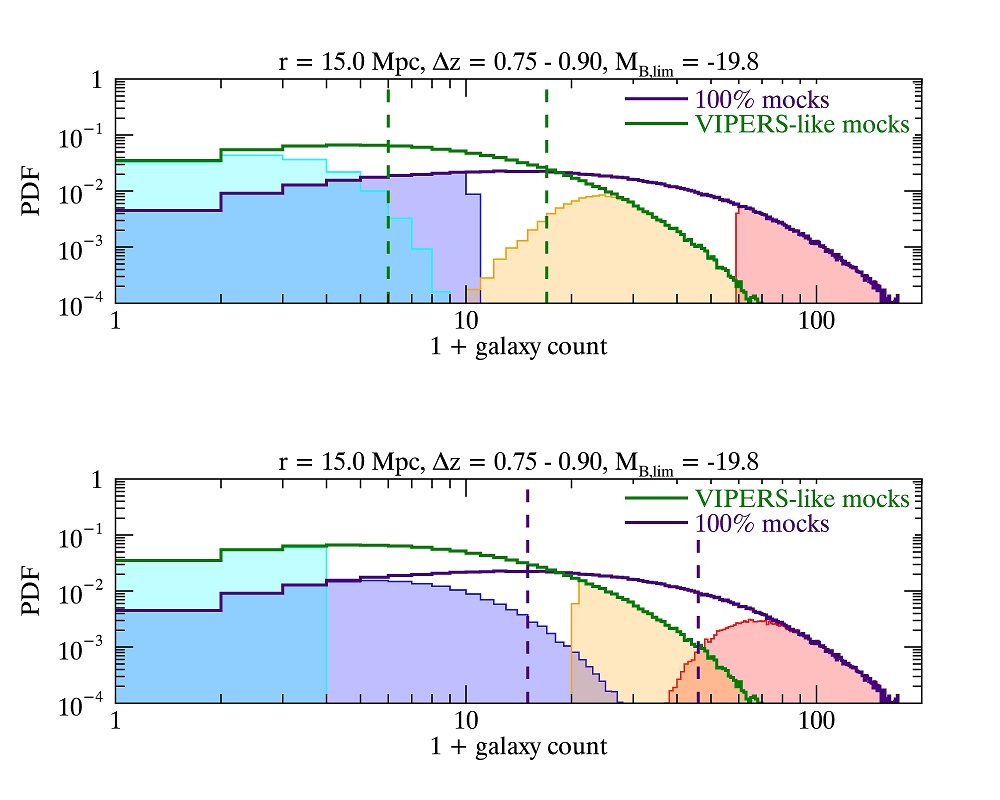}
\caption{Counts in cells distributions in the mock samples in redshift
  bin $0.75 < z <0.9$ for cells of radius $r = 15$ Mpc. The violet
  line shows the distribution for the 100\% samples and the green line
  for the VIPERS-like samples.  Top panel: cells selected from the
  100\% sample; blue and red shaded regions represent the 15\% least
  and most populated cells selected in the 100\% samples respectively;
  cyan and orange shaded regions show the distribution of cells when
  viewed in the VIPERS-like samples. The vertical dashed line
  indicates the 25th and 75th percentiles of the VIPERS-like sample
  distribution.  Bottom panel: vice versa, cells are selected from the
  VIPERS-like sample and viewed in the 100\% sample.  The vertical
  dashed line indicates the 25th and 75th percentiles of the 100\%
  sample distribution.}
\label{UMccount}
\end{figure}

How does the survey selection function, including the sampling rate
and mask, affect the detection of large-scale structures? To evaluate
the impact of observational biases, we compare the distribution of
counts-in-cells measured in the VIPERS-like mock sample with those
measured in the 100\% mock sample.  We do not expect an important
contribution from the cross-shaped unobserved regions or from the
missing quadrants, since their dimensions are smaller than the spheres
we have used for the counts-in-cells (see Fig. \ref{VIMOSgaps}). Since
the comoving size of the gaps is highest in the farthest redshift bin,
we expect any effect to be more relevant to small cells at higher
redshift.  Fig. \ref{UMccount} presents histograms for counts-in-cells
for cell radius $r = 15$ Mpc and redshift bin $0.75 < z < 0.9$, for
cells selected from the 100\% and viewed in the VIPERS-like mock (and
vice versa).

As for the previous tests, we see that more than 95\% of the cells at
the extremities of the distributions remain there after transitioning
between the two samples. Also, we see that the PDF of the 100\% mock
sample is more skewed than that of the VIPERS-like mock, because of
the reduced sampling rate. The under-sampling of the VIPERS-like mock
smooths the density contrast seen in the 100\% mock. The sampling rate
effects dominate over the redshift-space effects discussed earlier.
However, we conclude that on the scales we are interested in, masking
and selection effects do not significantly introduce artificially
underdense regions.

\subsection{Sampled volume} \label{svol}
The dimensions of the survey fields place limits on the scale of under
densities that may be found. The comoving transverse sizes of the W1
and W4 fields at different redshifts are presented in table
\ref{trasize}. We note that the minimum dimension of the fields are a
factor of $\sim4$ times greater than the scale of the voids of
interest ($\gtrsim15$Mpc), over most of the redshift range.
\begin{table}[!h]
\centering
\caption[The comoving transverse size of W1 and W4 samples along RA
  and Dec directions at different redshift.]{Comoving transverse size
  of W1 and W4 samples.}
\label{trasize}
\begin{tabular}{ccc}
\hline
\noalign{\smallskip}
z & W1 (Mpc) & W4 (Mpc) \\
\hline
\noalign{\smallskip}
0.55 & 308 $\times$ 47.1 & 188 $\times$ 54.3 \\
\noalign{\smallskip}
0.75 & 399 $\times$ 61.0 & 244 $\times$ 70.4 \\
\noalign{\smallskip}
0.9 & 461 $\times$ 70.5 & 282 $\times$ 81.4 \\
\hline
\end{tabular}
\end{table}

We consider the volume of the survey fields in which we may place a
spherical cell such that greater than 80\% of the cell volume falls
within the survey boundary.  We measure this survey effective volume by
randomly placing spheres of four different cell radii, between 15 and
30 Mpc, inside the survey and counting the percentage of spheres that
meet the volume requirement, see table \ref{svolfra}.

We see that the effective volume decreases for the largest sphere
sizes. Thus, we expect there to be a non-negligible selection effect
on the size of the observed spheres as a function of redshift. The
larger spheres whose centres are not positioned within the survey effective
volume will be detected as multiple smaller spheres, further modifying
the size distribution. We will discuss the impact of these effects on
our measurements in the next section.

\begin{table}
\centering
\caption[The table presents the volume fraction of W1 and W4
  volume-limited samples that is effectively sampled when searching
  for empty spheres of a given size.]{Sampled volume fraction.}
\label{svolfra}
\begin{tabular}{ccc}
\hline
\noalign{\smallskip}
\multicolumn{3}{c}{W1}\\
\noalign{\smallskip}
r (Mpc) & z = 0.55 - 0.75 & z = 0.75 - 0.9 \\
\hline
\noalign{\smallskip}
15 & 69.5 \% & 74.6 \% \\
\noalign{\smallskip}
20 & 60.3 \% & 66.8 \% \\
\noalign{\smallskip}
25 & 50.5 \% & 59.3 \% \\
\noalign{\smallskip}
30 & 40.5 \% & 51.5 \% \\
\hline
\hline
\noalign{\smallskip}
\multicolumn{3}{c}{W4}\\
\noalign{\smallskip}
r (Mpc) & z = 0.55 - 0.75 & z = 0.75 - 0.9 \\
\hline
\noalign{\smallskip}
15 & 73.6 \% & 78.1 \% \\
\noalign{\smallskip}
20 & 65.4 \% & 71.2 \% \\
\noalign{\smallskip}
25 & 57.4 \% & 65.0 \% \\
\noalign{\smallskip}
30 & 48.7 \% & 57.8 \% \\
\hline
\end{tabular}
\end{table}

\section{Void detection in VIPERS}
The algorithm presented in this paper is based on the detection of
empty spheres, and the voids are defined as regions devoid of galaxies
with absolute magnitude brighter than a specified limit ($M_B = -19.8$
evolving with redshift). The use of spheres is a simple approximation
that allows us to easily locate regions of interest. The main
disadvantage is that a single sphere will not be sufficient to
reconstruct the real shape of the examined empty region. Yet using
spheres does not restrict our ability to detect voids of non-spherical
shape. Simply, more than one sphere will be found inside the same
cosmic void. Furthermore, a simple percolation analysis may be used to
define a posteriori the volumes that correspond to topologically
connected spheres satisfying our density cut-off (see section
\ref{voidperc}).

\subsection{Isolated and unisolated galaxies}\label{isolate}
First, the galaxies in the sample are classified as isolated or
unisolated galaxies, in a similar fashion to the methods of
\citet{eladpiran97} and \citet{hoylevogeley02}. The idea is that, in
the large-scale distribution of matter, the voids are surrounded by
structures with high galaxy density (walls and filaments). The
low-density regions, however, are not totally empty. They contain both
void galaxies and spurious galaxies introduced to the void by redshift
space distortions. Therefore, it is necessary to excise the galaxies
that would cause these regions to be split into smaller volumes. Care
must be taken for galaxies near the survey borders which could
appear as isolated because their actual neighbouring galaxies are
outside the survey limits.

We identify and remove isolated galaxies using an iterative
procedure. In the algorithm, isolated galaxies are defined as galaxies
far away from the survey borders with fewer than three neighbouring
galaxies within a comoving sphere of radius, $l_{lim}$, where
$l_{lim}$ is defined from the galaxy distribution itself.

The steps are as follows.
\begin{enumerate}
\item The distribution of the third nearest neighbour distance is
  computed for unisolated galaxies (initially all galaxies are
  classified as unisolated).
\item The limit radius is calculated from this distribution. It is defined as:
\begin{equation}
\centering
l_{lim} = \bar{l}_{3NN} + 2.5 \sigma_{3NN} ,
\end{equation}
where $\bar{l}_{3NN}$ and $\sigma_{3NN}$ are the mean and the standard
deviation of the distribution of distances to the third nearest
neighbour.
\item Galaxies that do not have three neighbours within the limit are
  classified as isolated unless a sphere of radius $l_{3NN}$, centred
  on the galaxy, is less than 80\% inside the survey borders.
\end{enumerate}
The steps converge quite rapidly and are repeated until no more
isolated galaxies are found. Isolated galaxies are then removed from
the sample leaving the cleaned sample. The number of isolated
galaxies identified in each of the two VIPERS fields is provided in
Table \ref{w1w4cleaning}.

\begin{table}
\centering
\caption[The table presents the number of galaxies contained in the
  volume-limited samples obtained from the real data, and the number
  of those identified as isolated.]{Detection of isolated galaxies in
  W1 and W4 samples.}
\label{w1w4cleaning}
\begin{tabular}{ccc}
\hline
\noalign{\smallskip}
Sample & Total galaxies & Isolated galaxies \\
\hline
\noalign{\smallskip}
W1 & 16249 & 931 \\
\noalign{\smallskip}
W4 & 12300 & 604 \\
\hline
\end{tabular}
\end{table}

\subsection{Empty spheres}
The next step of the algorithm is the search for empty spheres in the
cleaned sample.  A three-dimensional grid of comoving step size
1 Mpc is super-imposed on the sample. We compute the distance from each
grid point to the nearest galaxy. This is the radius of the largest
empty sphere centred on that point.  We then calculate the volume
fraction inside the survey boundaries for each sphere. Spheres with a
volume fraction inside the borders lower than 80\% are rejected and
not included in the further analysis.

\subsection{Maximal spheres}\label{maximal}
\begin{figure} 
\centering
\includegraphics[width=7cm]{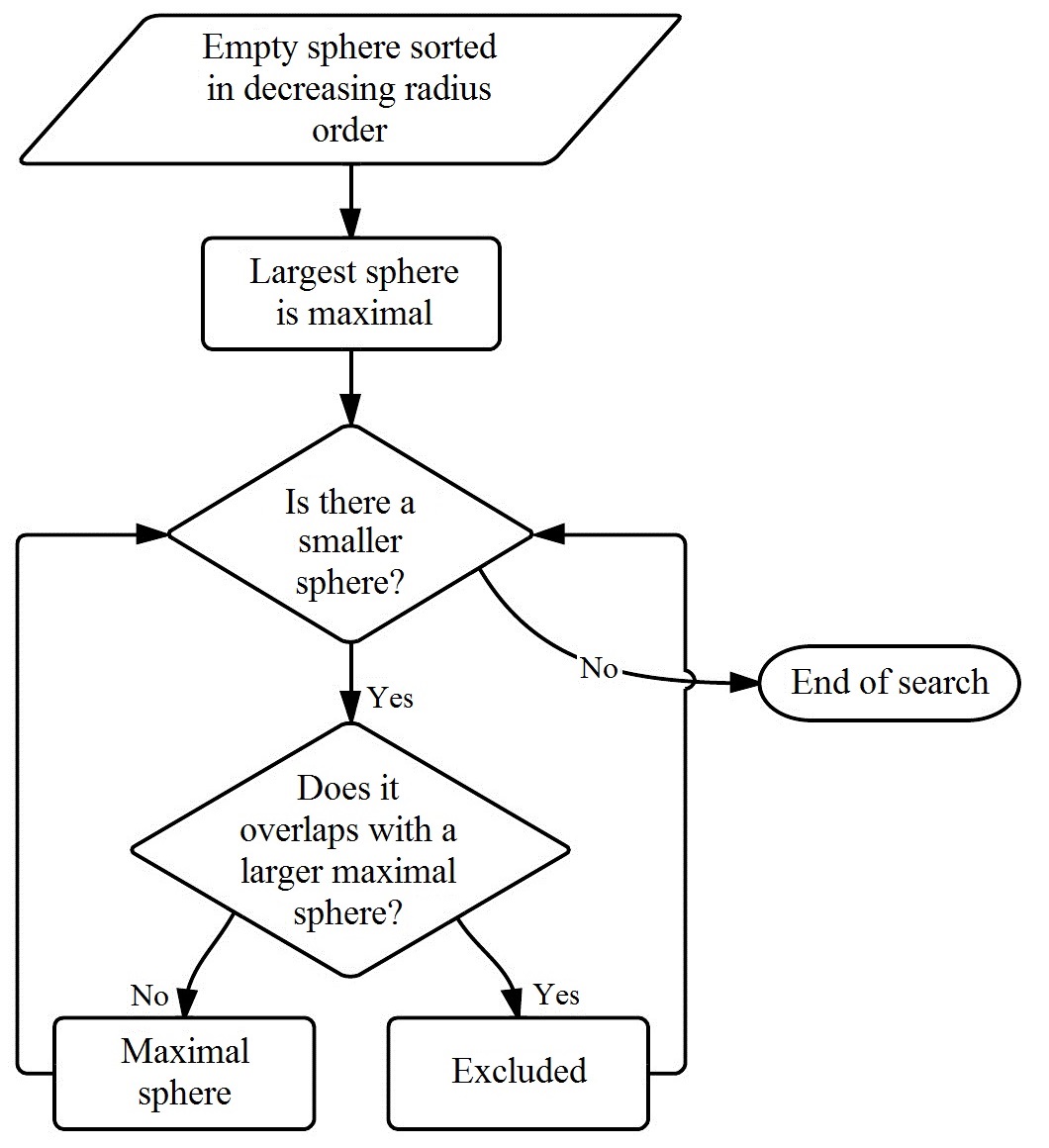}
\caption{Maximal spheres identification process.}
\label{maximal_flow}
\end{figure}

The spheres are sorted by size. The largest sphere is defined as a
maximal sphere, we then check to see if the next largest sphere
overlaps with this one, if it does not then it is also defined as
being a maximal sphere. We continue down the list of spheres, checking
for overlap with all maximal spheres already detected, until all
spheres have been counted (see Fig. \ref{maximal_flow}).

\subsection{Statistical significance} \label{ADstatsig}

To assess the statistical significance of maximal spheres, we compare
the occurrence of these spheres to the occurrence of maximal spheres
in Poisson-distributed catalogues with the same number density and
mask as VIPERS.  The confidence level with which we detect a maximal
sphere in the galaxy distribution (over the Poisson distribution) is
then given by
\begin{equation}
P(r) = 1 - \frac{N_{rnd}(\geq r)}{N_{rnd,tot}} ,
\end{equation}
where $N_{rnd}(\geq r)$ is the number of maximal spheres found in the
random sample with radius equal to or greater than $r$, and
$N_{rnd,tot}$ is the total number of maximal spheres in the random
sample. The closer $P(r)$ is to 1, the less likely a maximal sphere is
expected to occur in a random distribution.

The number density of galaxies in the VIPERS data is not constant with
redshift (see Fig. \ref{dsep1}). Therefore, the significance limit for
the radius of empty spheres will also depend on redshift. To calculate
this limit we have used random catalogues with different density
values. The density values correspond to the mean density values of
our VIPERS fields as a function of redshift

\begin{figure} 
\centering
\includegraphics[width=5cm]{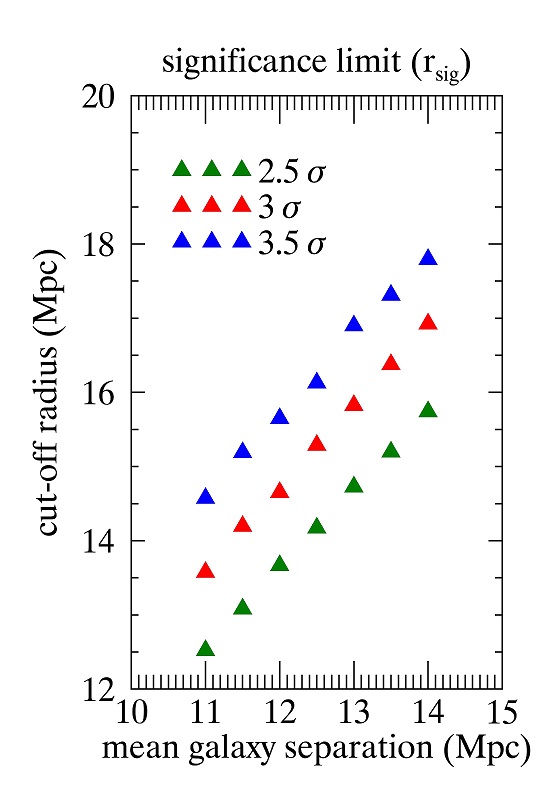}
\caption{Significant radius limit for maximal spheres as a function of
  the mean inter-galaxy separation for confidence levels 2, 3, and 3.5
  $\sigma$.}
\label{r_sig}
\end{figure} 

Fig. \ref{r_sig} shows the significance limit for maximal spheres with
varying mean inter-galaxy separation for different confidence
levels. As can be seen, the trend is quite linear. The relationship
between the cut-off radius, corresponding to a $3 \sigma$ significance
for the maximal spheres, and the mean inter-galaxy separation can be
described with the following fitting formula:
\begin{equation}
r_{cut-off} = 1.35 + 1.11\,d_{sep}.  
\end{equation}
Using this fit, and the trend of $d_{sep}$ with redshift, we find the
trend of the cut-off radius with redshift.  The trend for $d_{sep}$ is
obtained by fitting the combined data from the two fields W1 and W4
where isolated galaxies have been removed (similar to what was shown
in Fig. \ref{dsep1}, where the isolated galaxies have not yet been
identified), and can be described as:
\begin{equation}
d_{sep} = 
\begin{cases}
11.95 & z \leq 0.775\\
15.07 z + 0.28 & z > 0.775
\end{cases}
\end{equation}
Combining the two previous equations, the trend of the cut-off radius
with redshift yields then the following values:
\begin{equation}
r_{cut-off} = 
\begin{cases}
14.62 & z \leq 0.775\\
16.73 z + 1.66 & z > 0.775 .
\end{cases}
\end{equation}

\begin{figure} 
\centering
\includegraphics[width=9cm]{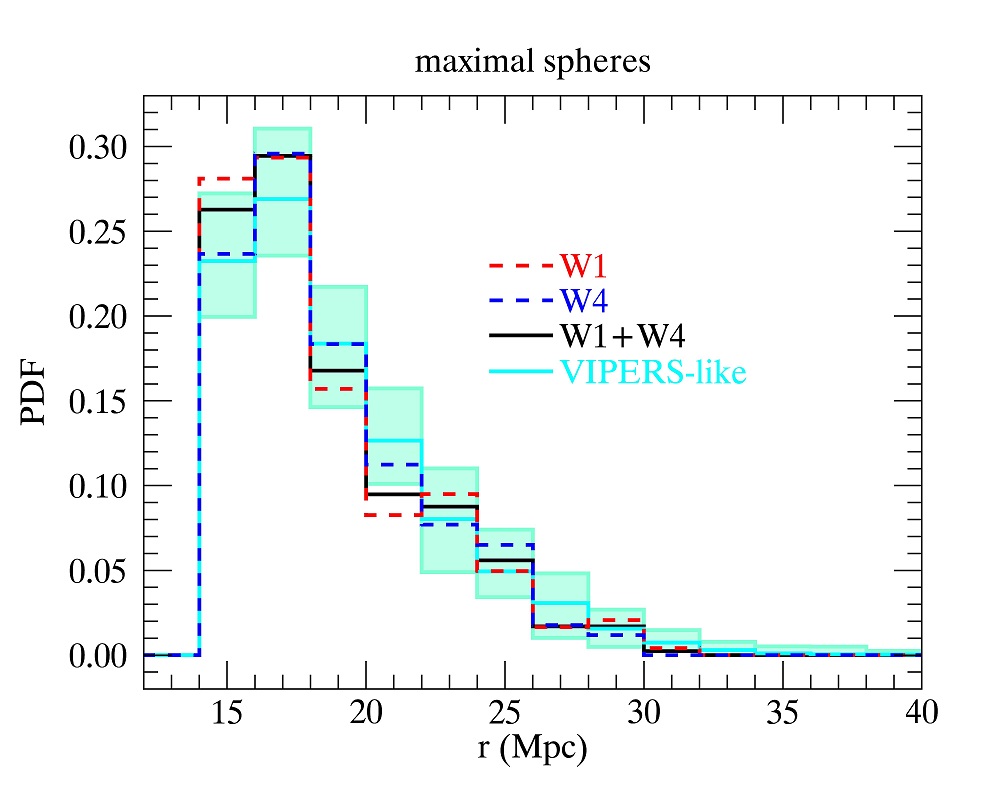}
\caption{Size distribution for maximal sphere found in W1 and W4 field
  samples and in VIPERS-like samples. The shaded regions correspond to
  the standard deviation of the measurements in the mock catalogues.}
\label{WMOCKmaxsphere}
\end{figure} 

Fig. \ref{WMOCKmaxsphere} shows the size distribution of the maximal
spheres detected in the VIPERS survey as compared to the VIPERS-like
mocks. It can be seen that the distributions obtained from the mock
samples are in good agreement with measurements from the data.

\begin{figure*} 
\centering
\includegraphics[width=0.45\textwidth]{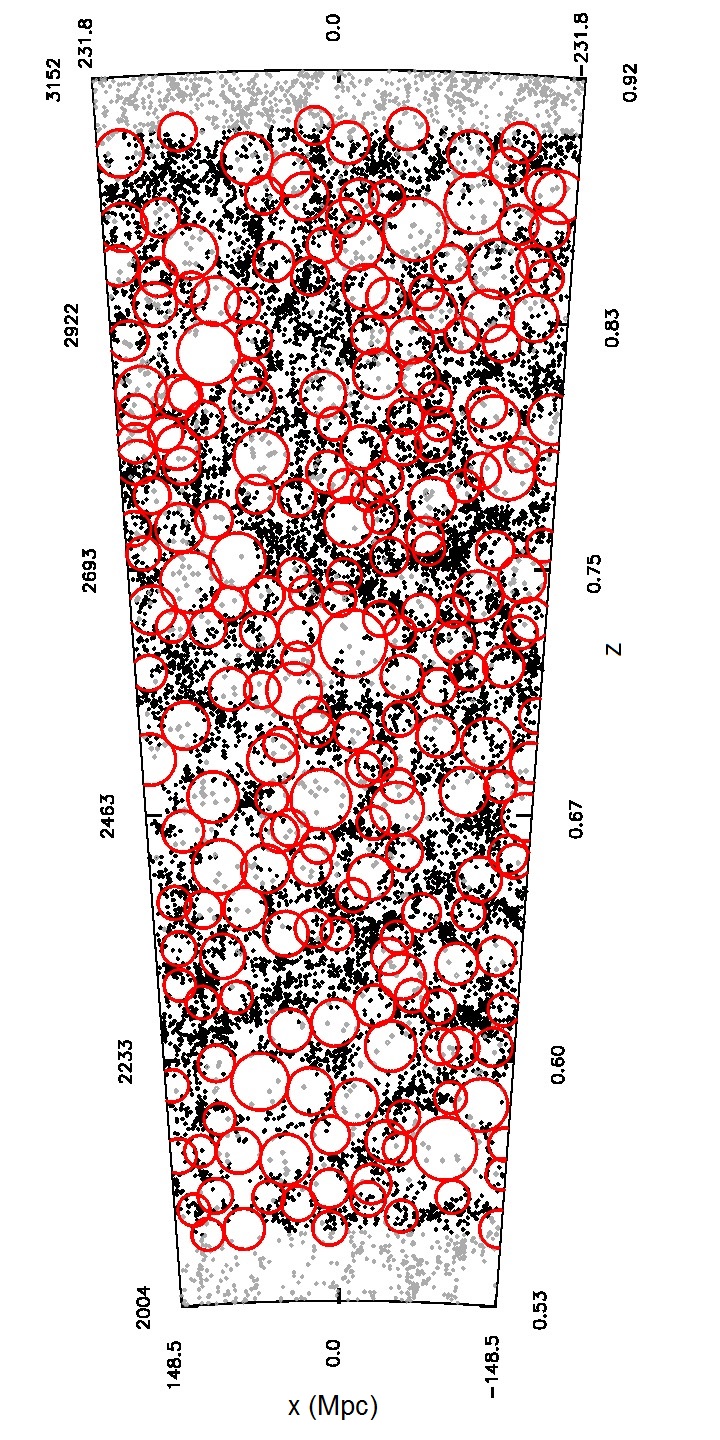}
\includegraphics[width=0.45\textwidth]{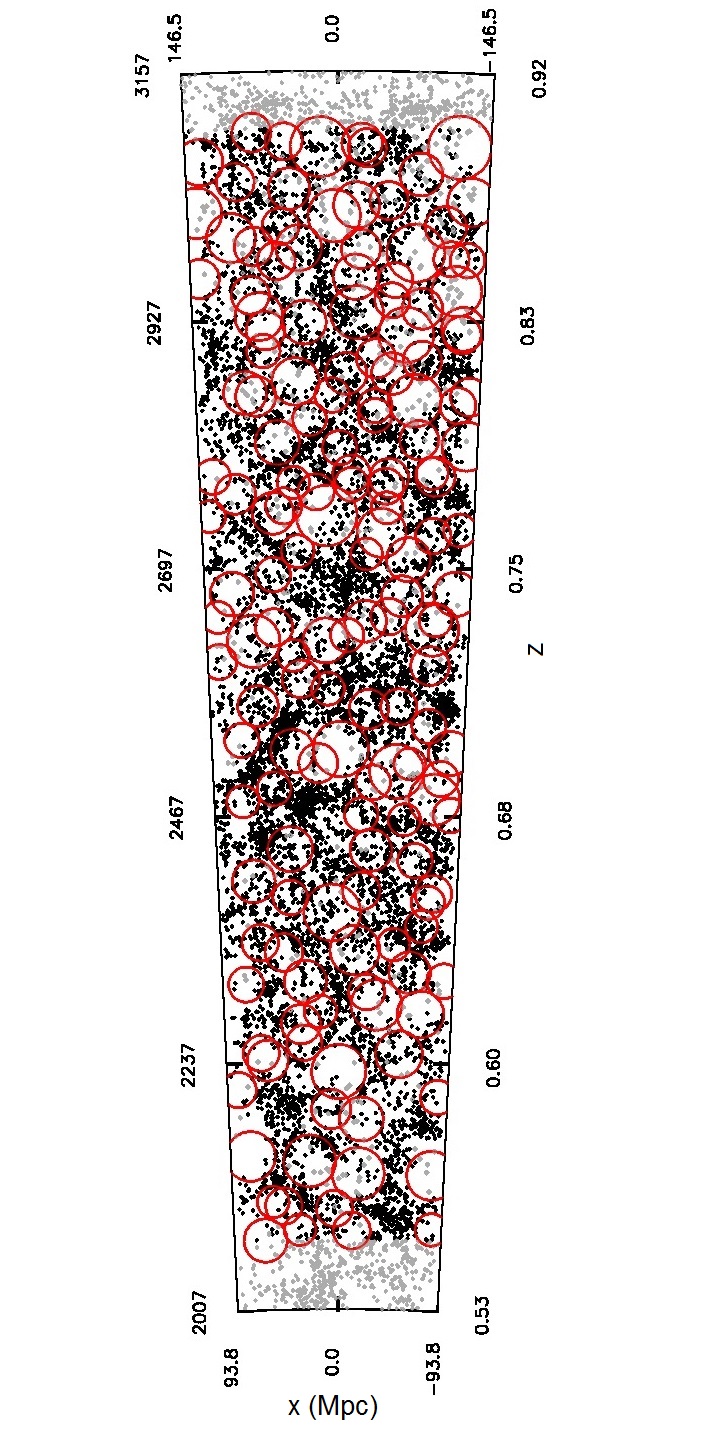}
\caption{Maximal spheres in the W1 (left) and W4 (right) samples; grey
  points are galaxies detected as isolated and galaxies outside the
  sample redshift range. The scales show comoving distance in Mpc and
  the corresponding redshift.}
\label{W1cone}
\end{figure*} 

Figure \ref{W1cone} shows the position of the maximal spheres in the
two VIPERS fields projected along the RA-redshift plane. These maximal
spheres correspond well to the appreciably low-density regions, as one
can see by eye. We found 229 maximal spheres in the W1 field and 159
in W4. Because fields are not so extended in declination, large
maximal spheres ($\geq 25 {\rm Mpc}$) can only be observed in a
fraction of the volume. Therefore, the number of larger spheres found
is probably an underestimate.

\subsection{Void percolation}\label{voidperc}
\begin{figure*} 
\centering
\includegraphics[width=9cm]{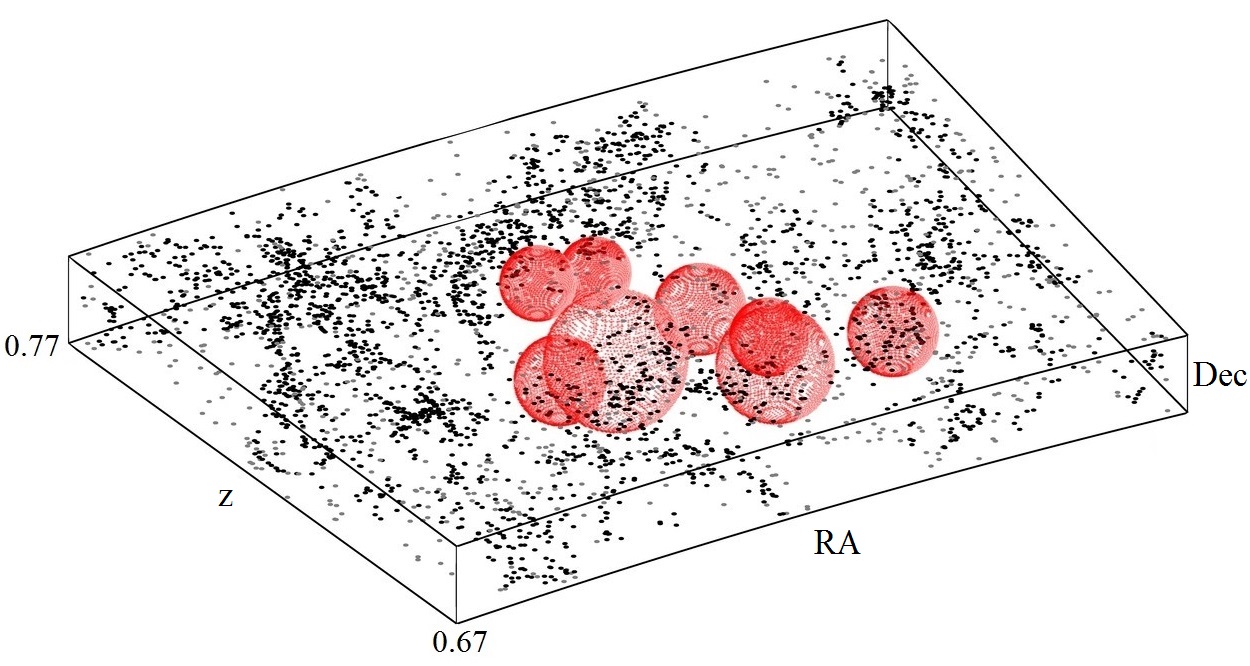}
\includegraphics[width=9cm]{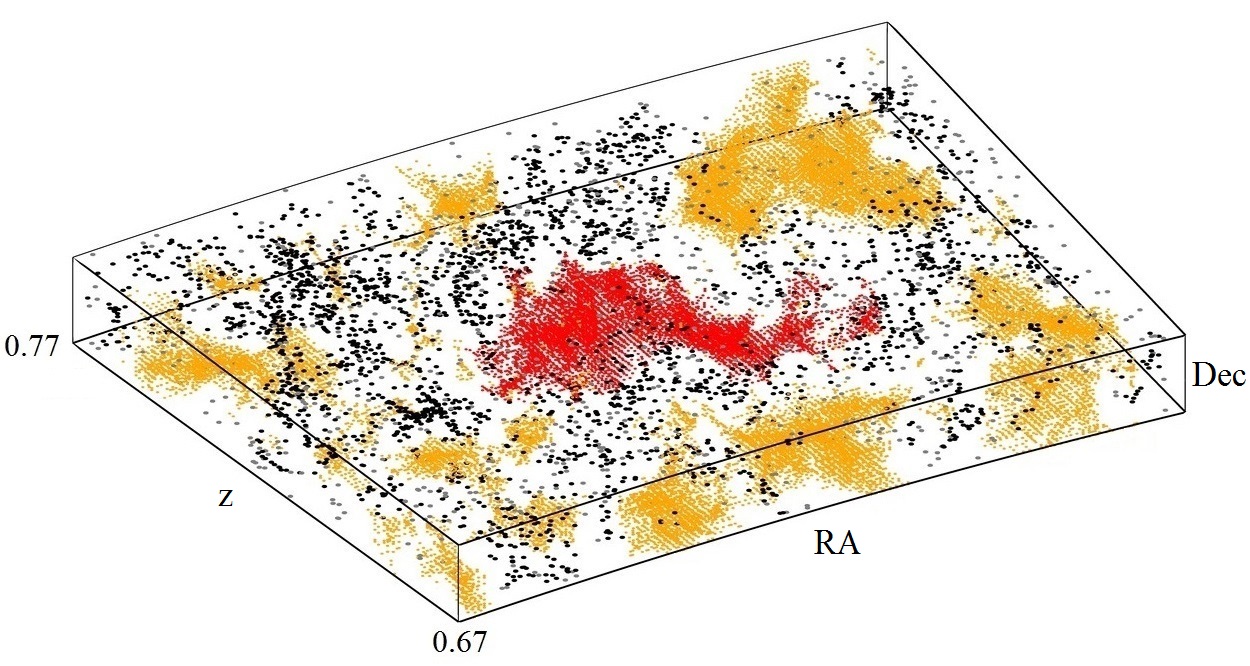}
\caption{Illustration of the region surrounding the largest maximal
  sphere in our catalogue (W1.0075.000), which has radius of 31
  Mpc. The left-hand panel shows this sphere and the other six maximal
  spheres detected in this void region. The right-hand panel shows, in
  red, the centres of the overlapping significant spheres that make up
  the void, other void regions within this area of the survey are
  shown in orange. The black points in both panels are the unisolated
  galaxies, while the grey points are the isolated galaxies.}
\label{fig:voidperc}
\end{figure*}

The identified maximal spheres may mark the centres of individual
voids or may join together to trace larger irregularly shaped
voids. To identify voids traced by multiple spheres, we link together
the grid centres where spheres with sizes greater than our significance
limit are centred (irrespective of their being maximal or not). Groups
of spheres are identified using a friends-of-friends algorithm with a
linking length equal to the grid step size. The groups of sphere
centres define the void regions that are connected, see
Fig. \ref{fig:voidperc}, thus demonstrating that our method can easily
be used to also define under-densities of complex shape. Void regions
defined in this manner are by definition topologically distinct
regions with an unisolated galaxy density lower than some threshold,
$\delta_g < 1/V(r_{cut-off}) - 1$, where $V(r_{cut-off})$ is the volume
of a sphere of radius $r_{cutoff}$. However, the set of maximal
spheres (defined in section \ref{maximal}) are sufficient for the
subsequent analysis presented in this paper.

\section{Void catalogue}
The catalogue of maximal spheres for both VIPERS fields is presented
in Table \ref{voidtable}. The spheres are linked into larger connected
voids as discussed Sec. \ref{voidperc}.

The identification number of each maximal sphere is composed of: the
VIPERS field to which the maximal sphere belongs ({\tt W1} or {\tt
  W4}); a void index indicating to which void the sphere belongs ({\tt
  VVVV}); a sphere index ({\tt SSSS}) indicating the size rank of the sphere
relative to other maximal spheres within the same void. The complete
void ID has the form {\tt Wx.VVVV.SSSS}.

For each maximal sphere of our catalogue we provide: the right
ascension, declination, and redshift of the sphere's centre; its
comoving radius in Mpc; a p-value giving the detection significance
with respect to a Poisson distribution. 

\section{Void-galaxy cross-correlation function} \label{ccorr}
The void-galaxy cross-correlation function measures the probability,
in excess of random, of finding a galaxy at a certain distance from
the centre of a void.  Cross correlations between galaxies and other
astronomical objects, such as groups, clusters, and quasars, have been
studied extensively \citep{padillaetal01, padillalambas03, Myers2003,
  Yang2005, Mountrichas2009, Knobel2012}. However, the study of
void-galaxy cross correlations is in its infancy.

Nevertheless, the void-galaxy cross-correlation function has the
potential to be a valuable statistic. It contains information that 
can be used to constrain models of galaxy bias
\citep{hamausetal13}. It might also be possible to use the geometrical
properties of the void-galaxy cross correlation function as a standard
ruler \citep{sutteretal12b}.

The void-galaxy cross correlation function contains information on the
mean density profile of the voids and on the dynamics of the tracer
population \citep{padillaetal05, pazetal13}. This information can be
used to discriminate between different theories of gravity
\citep{martinsheth09}.

There is an on-going debate in the literature on the universality of
void density profiles \citep{hamausetal13, 2014MNRAS.440..601R,
  2014MNRAS.440.1248N, 2014arXiv1407.1295N}. It is generally agreed
that void density profiles can be divided broadly into two categories:
compensated and uncompensated voids. Compensated voids are surrounded
by an over-dense shell and may indeed be embedded in over-dense regions
that are eventually going to collapse, destroying these
voids. Uncompensated voids are not surrounded by an over-dense shell
and represent under-dense regions that will continue to exist in
the future \citep{2004MNRAS.350..517S}. Our decision to include only
the most significantly empty spheres in our catalogue means that our
voids are relatively large. Generally they should correspond to
uncompensated voids and so we should not expect to see a strong ridge
in the correlation function.

The centres of the maximal spheres do not represent the centres of
spherical under densities. However, in principal, there should be no
preferred direction to the asymmetry of the voids in our
catalogue. Therefore, stacking the maximal spheres should produce an
axially symmetric density profile.

Here we search for evidence of anisotropy in the void-galaxy
cross-correlation function. As previously mentioned, galaxies within
voids are expected to flow towards the edge of the void under the
influence of gravity. Since the redshift of a galaxy, a measure of its
recessional velocity, is used as a proxy for distance, these linear
outflows are expected to produce an enhancement of the void-galaxy
cross-correlation function in the line of sight direction relative to
the tangential direction. There should also be an apparent stretching
of voids along the line of sight. However, uncertainty in defining the
void centres will smooth out this effect. 

We measured the cross-correlation function using the \citet{davispeebles83} estimator,
\begin{equation}
\xi_{vg}(\eta,\alpha) = \frac{N_R}{N_g} \frac{DD(\eta,\alpha)}{DR(\eta,\alpha)} - 1 ,
\end{equation}
where DD($\eta,\alpha$) and DR($\eta,\alpha$) are the number of
void-galaxy and void-random pair counts as a function of void-galaxy
separation. $N_R$ and $N_g$ are the number of random points and the
number of galaxies, respectively. The coordinates $\eta$ and $\alpha$
are the void-galaxy radial separation (normalised to the radius of the
void), $\eta = r/R_v$, and the angle between the line of sight and the
line connecting the void centre with the galaxy.

The random catalogue has the same volume and angular selection
function as the real data. In order to have the correct radial
selection function, redshift values were randomly assigned from the
redshift values of galaxies in the VIPERS-like mocks.

Other estimators, such as the Landy-Szalay estimator, are less biased
than the estimator we are using, however, they require random
catalogues for both sets of objects being cross-correlated. Since we
cannot estimate the selection function of the voids in advance,
constructing a corresponding random catalogue is not
possible. Therefore, we will use the Davis and Peebles estimator.

We first measured the cross-correlation function in the mock
catalogues. To maximise the total number of void-galaxy pairs we
reintroduced the isolated galaxies (see section \ref{isolate}) and
used all the maximal spheres, including spheres close to the survey
borders. This ensures that we are including all available information
about the density profile of the voids.

\begin{figure*}[!ht] 
\centering
\includegraphics[width=9cm]{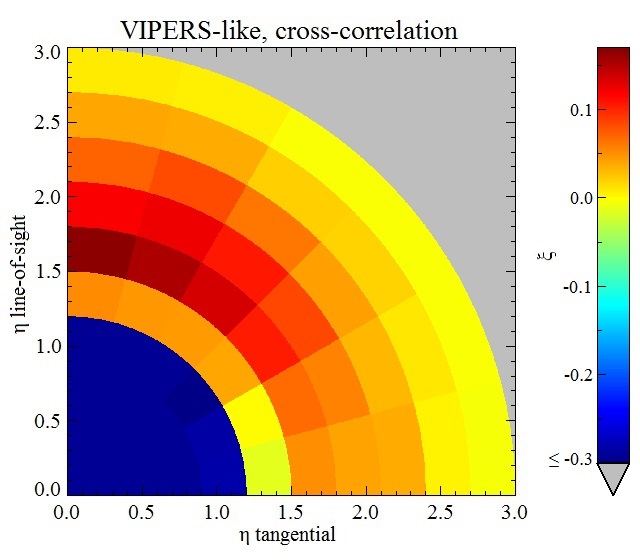}
\includegraphics[width=9cm]{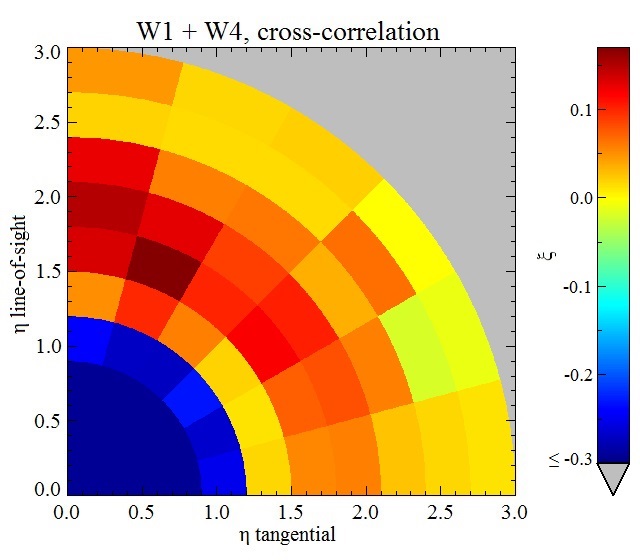}
\caption{Void-galaxy cross-correlation function, as measured in
    the mock catalogues (left panel) and in the VIPERS data (right
    panel). Measurements were made in ten radial bins and in six
  angular bins. The axes plotted correspond to the tangential,
  $\theta$, and line of sight, $\pi$, directions. The enhancement in
  the line of sight direction, visible in both the mock catalogues and
  in the data, is evidence of redshift space distortions caused by
  linear outflows.}
\label{fig:ccorr}
\end{figure*}

The mean void-galaxy cross-correlation from the mock catalogues is
plotted in the right-hand panel of figure \ref{fig:ccorr}. One can see
that close to the origin the cross-correlation function is close to
$\xi \sim -1$, rising to zero far away from the void centre. There is
also a clear anisotropy visible. The correlation function is enhanced
in the line of sight direction, peaking strongly between one and two
times the radius of the maximal spheres. This is the result of linear
redshift space distortions.

We then proceeded to measure the void-galaxy cross-correlation
function in the VIPERS data, illustrated in the left-hand panel of
Fig. \ref{fig:ccorr}. Similarly to the mocks, there is a clear
enhancement of the correlation function in the line of sight
direction.

Using the variance of the measurements of the mock catalogues we are
able to calculate the $\chi^2$ between our measurement of
$\xi_{vg}(\eta,\alpha)$ in the VIPERS data and in the mocks. The value
quantifies the agreement between the data and mock catalogues.  The
computation of the $\chi^2$ requires knowing the inverse of the
covariance matrix of data points.  In our case, we cannot estimate it
with sufficient accuracy with the mock catalogues and so we use only
the variance neglecting the covariance terms.  We find that the
reduced $\chi^2 = 0.49$ (per degree of freedom, for the 60
$\eta$-$\alpha$ bins in Figure \ref{fig:ccorr}), which is a very good
fit.  This value may be lower than expected since we have neglected
the correlations between data points.  This supports the validity of
the concordance cosmology and the halo model used to generate the mock
catalogues.

\begin{figure}[!h] 
\centering
\includegraphics[width=9cm]{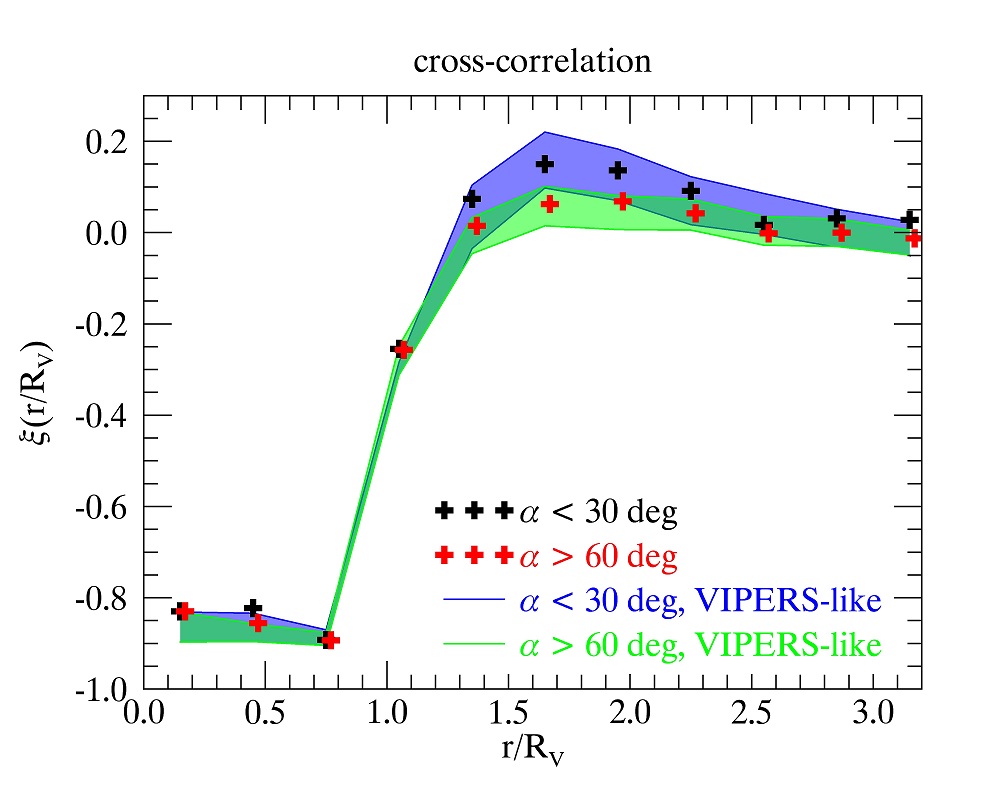}
\caption{The angle-average void-galaxy correlation function as a
  function of radial distance normalised to the void radius. To
  demonstrate the anisotropy we average over two angular wedges, along
  the line of sight ($\alpha<30\degr$, black line) and transverse to
  the line of sight ($\alpha>60\degr$, red line). The shaded regions
  represent the spread of values measured in the mocks. The
  enhancement along the line-of-sight is an indication of the
  redshift-space distortion produced by the outflow of galaxies from
  voids.}
\label{fig:ccorr_r}
\end{figure}

To highlight the enhancement along the line of sight we have plotted
in Figure \ref{fig:ccorr_r} the VIPERS void-galaxy cross-correlation
function as a function of radial separation for two angular bins, one
close to the line of sight, $\alpha < 30 \degr$, and one close to the
plane of the sky, $\alpha > 60\degr$. One can see that, particularly
in the range 1-2.5 void radii, the line of sight cross-correlation
function is greater than the parallel. Furthermore, for both the
$\alpha < 30 \degr$ and $\alpha > 60\degr$ cases, the
cross-correlation lies within the range of values measured in the
mocks. This further demonstrates the good agreement between the data
and the mock catalogues.

\section{Discussion and conclusions}
VIPERS (VIMOS Public Extragalactic Redshift Survey;
\citealp{guzzoetal13}) is mapping the large-scale distribution of
galaxies at redshift $0.5<z<1.2$ and provides a unique volume in which
to study the distribution of voids in the galaxy distribution at
moderate redshift. 

The identification of voids in the galaxy distribution is challenging
and it is made more difficult by observational systematics. These
effects are particularly important for VIPERS, which has a complex
geometry including internal gaps. The two VIPERS fields, W1 and W4,
have transverse comoving dimensions of $\sim70\times350$ Mpc (see
Table \ref{trasize}) and the narrow dimension further limits the
volume in which we may identify large under-densities. The sampling
rate is 35\% to an apparent flux limit of $i_{AB}=22.5$
\citep{guzzoetal13}. In addition, the survey strategy leaves gaps in
the sky coverage (See Fig. \ref{w1w4borders}). Using counts-in-cell
measurements on mock catalogues we investigated how observational
systematics and redshift-space distortions modify the true density
field. We find that on scales of $r \gtrsim 15$Mpc the tails of the
counts-in-cells PDF are well preserved such that we can truly identify
the emptiest regions of the survey.

Void search methods, such as those using Voronoi tessellation
\citep{Platenetal2007, 2008MNRAS.386.2101N, sutteretal12}, are
unsuitable for this particular survey because of careful corrections
required for borders and gaps. Other considerations, such as the lack
of breadth in declination coverage and the limitations of the VIPERS
survey strategy, make it difficult to apply void detection methods
such as the water-shed method that require contiguous volumes.

In this paper, we have presented a general void-search algorithm
capable of finding empty regions in a galaxy redshift survey such as
VIPERS with irregular borders and internal gaps. The method is based
on the identification of spheres that fit between galaxies. We show
that the voids may be well characterised by keeping only the
significant spheres, those that are unlikely to be found in a uniform
Poisson distribution with the same number density. The significance
limit for VIPERS gives voids with radii greater than
$\sim$15Mpc. These spheres trace empty regions of arbitrary shape, as
shown in Fig. \ref{fig:voidperc}.

The set of largest spheres that do not overlap are termed
maximal spheres and we find 411 maximal spheres in the VIPERS
survey with radii $r\gtrsim15$Mpc between $0.55<z<0.90$. The
properties of this special subset may be used to characterise the void
distribution.

We present a catalogue of maximal spheres identified in VIPERS. The
spheres have been associated with larger under-densities using the
percolation method, thus we associate each sphere with a void ID
indicating larger empty regions.

We note that the identification of the maximal spheres is affected by
the survey geometry. We see that the effective volume decreases for
the largest sphere sizes (Table \ref{svolfra}). Thus, we expect there
to be a noticeable selection effect. The larger spheres whose centres
are not positioned within the effective volume will be detected as
multiple smaller spheres, since the effective volume is larger for
smaller spheres.  This will affect our ability to compute some void
statistics such as the volume filling factor as a function of
redshift, since the splitting of large spheres that are not found has
a different impact at different redshifts. The smaller spheres could
also fall below the significance limit (see Sec. \ref{ADstatsig}),
which means a loss of detected volume.

Our selection biases prevent us from drawing conclusions from the actual
void size distribution, but we can directly compare the void size
distribution as observed in the VIPERS data with that observed in the
mocks, as both are affected by the same selection biases.

Using our catalogue of maximal spheres, we compute the void-galaxy
cross-correlation function. We find an enhancement in the correlation
function along the line of sight. This anisotropy is a clear signature
of the velocity flows induced by the voids and matches well the signal
measured in mock catalogues.

In the next phase of this work, we will fit the mean density profile
of the voids and model the void-galaxy cross-correlation function
(Hawken et al. in prep). With constraints on the galaxy bias, this
analysis will provide a measurement of the distortion, which may be
related directly to the linear growth rate of structure, $d\log
\delta/d\log a$. We may further characterise the distribution of voids
through topological analyses. For example, the Minkowski functionals
may be measured directly on the spheres complementing works on the
topology of the galaxy distribution (Schimd et al, in
prep.). Additionally, the isolated galaxies identified in this work can
be used to study galaxy formation in low-density environments.  With
the detection of cosmic voids in VIPERS, we open the door to new
cosmological constraints and detailed topological measurements of the
distant universe.

\section*{Acknowledgements}
This work is based on observations collected at the European Southern
Observatory, Cerro Paranal, Chile, using the Very Large Telescope
under programs 182.A-0886 and partly 070.A-9007.  Also based on
observations obtained with MegaPrime/MegaCam, a joint project of CFHT
and CEA/DAPNIA, at the Canada-France-Hawaii Telescope (CFHT), which is
operated by the National Research Council (NRC) of Canada, the
Institut National des Sciences de l'Univers of the Centre National de
la Recherche Scientifique (CNRS) of France, and the University of
Hawaii. This work is based in part on data products produced at
TERAPIX and the Canadian Astronomy Data Centre as part of the
Canada-France-Hawaii Telescope Legacy Survey, a collaborative project
of NRC and CNRS. The VIPERS web site is http://www.vipers.inaf.it/.
We acknowledge the crucial contribution of the ESO staff for the
management of service observations. In particular, we are deeply
grateful to M. Hilker for his constant help and support of this
program. Italian participation to VIPERS has been funded by INAF
through PRIN 2008 and 2010 programs. DM gratefully acknowledges
financial support of INAF-OABrera. LG, AJH, and BRG acknowledge
support of the European Research Council through the Darklight ERC
Advanced Research Grant (\# 291521).  AP, KM, and JK have been
supported by the National Science Centre (grants
UMO-2012/07/B/ST9/04425 and UMO-2013/09/D/ST9/04030), the Polish-Swiss
Astro Project (co-financed by a grant from Switzerland, through the
Swiss Contribution to the enlarged European Union), and the European
Associated Laboratory Astrophysics Poland-France HECOLS.  KM was
supported by the Strategic Young Researcher Overseas Visits Program
for Accelerating Brain Circulation (\# R2405).  OLF acknowledges
support of the European Research Council through the EARLY ERC
Advanced Research Grant (\# 268107).  GDL acknowledges financial
support from the European Research Council under the European
Community's Seventh Framework Programme (FP7/2007-2013)/ERC grant
agreement \# 202781. WJP and RT acknowledge financial support from the
European Research Council under the European Community's Seventh
Framework Programme (FP7/2007-2013)/ERC grant agreement \# 202686. WJP
is also grateful for support from the UK Science and Technology
Facilities Council through the grant ST/I001204/1. EB, FM and LM
acknowledge the support from grants ASI-INAF I/023/12/0 and PRIN MIUR
2010-2011. LM also acknowledges financial support from PRIN INAF
2012. YM acknowledges support from CNRS/INSU (Institut National des
Sciences de l'Univers) and the Programme National Galaxies et
Cosmologie (PNCG). CM is grateful for support from specific project
funding of the {\it Institut Universitaire de France} and the LABEX
OCEVU.

\bibliographystyle{aa}
\bibliography{paper}

\appendix
\section{Void Catalogue}
\onecolumn
\begin{longtab}
\begin{longtable}{cccccr}
\caption{\label{voidtable} Maximal spheres detected in VIPERS samples, divided in groups corresponding to individual voids. 
Voids are sorted in ascending redshift order, with the redshift associated with each void being that of its largest sphere.
Spheres inside each void are sorted in decreasing radius order.}\\
\hline
Sphere ID & r & RA & Dec & z & p-value \\
\hline
\endfirsthead
\caption{Continued.} \\
\hline
Sphere ID & r & RA & Dec & z & p-value \\
\hline
\endhead
\hline
\endfoot
\hline
\endlastfoot
W1.0001.000 & 14.99 & 37.8618 & -5.0805 & 0.5500 & 1.06$\times 10^{-3}$ \\
W1.0002.000 & 15.91 & 34.7120 & -4.3632 & 0.5508 & 1.27$\times 10^{-4}$ \\
W1.0003.000 & 19.15 & 36.9180 & -4.8320 & 0.5511 & 4.49$\times 10^{-9}$ \\
W1.0004.000 & 17.75 & 30.3598 & -4.9144 & 0.5522 & 5.24$\times 10^{-7}$ \\
W1.0005.000 & 15.06 & 32.8733 & -4.7494 & 0.5545 & 9.03$\times 10^{-4}$ \\
W1.0006.000 & 14.94 & 38.1792 & -4.3673 & 0.5573 & 1.17$\times 10^{-3}$ \\
W1.0007.000 & 16.10 & 33.7292 & -5.2396 & 0.5592 & 7.63$\times 10^{-5}$ \\
W1.0008.000 & 14.91 & 36.2588 & -4.7768 & 0.5601 & 1.25$\times 10^{-3}$ \\
W1.0009.000 & 16.17 & 37.6147 & -4.9668 & 0.5610 & 6.43$\times 10^{-5}$ \\
W1.0010.000 & 18.12 & 34.7082 & -4.6684 & 0.5622 & 1.48$\times 10^{-7}$ \\
W1.0010.001 & 16.12 & 35.5006 & -5.1586 & 0.5580 & 7.41$\times 10^{-5}$ \\
W1.0011.000 & 18.42 & 33.6532 & -4.3712 & 0.5636 & 5.39$\times 10^{-8}$ \\
W1.0012.000 & 14.76 & 30.3783 & -5.1528 & 0.5680 & 1.64$\times 10^{-3}$ \\
W1.0013.000 & 16.34 & 38.4660 & -5.1499 & 0.5732 & 4.04$\times 10^{-5}$ \\
W1.0014.000 & 14.77 & 37.9259 & -4.5107 & 0.5743 & 1.61$\times 10^{-3}$ \\
W1.0015.000 & 29.52 & 31.8305 & -4.8300 & 0.5745 & 2.96$\times 10^{-21}$ \\
W1.0015.001 & 24.71 & 30.9943 & -4.4637 & 0.5878 & 3.68$\times 10^{-16}$ \\
W1.0015.002 & 19.22 & 33.2973 & -4.3787 & 0.5760 & 3.44$\times 10^{-9}$ \\
W1.0015.003 & 15.53 & 31.5837 & -5.1299 & 0.5606 & 3.24$\times 10^{-4}$ \\
W1.0015.004 & 14.93 & 32.9740 & -5.2293 & 0.5739 & 1.18$\times 10^{-3}$ \\
W1.0016.000 & 15.46 & 30.3522 & -4.6177 & 0.5767 & 3.82$\times 10^{-4}$ \\
W1.0017.000 & 14.69 & 37.4070 & -4.9605 & 0.5833 & 1.88$\times 10^{-3}$ \\
W1.0018.000 & 15.19 & 32.8899 & -5.1968 & 0.5833 & 6.94$\times 10^{-4}$ \\
W1.0019.000 & 14.92 & 31.7586 & -5.2191 & 0.5895 & 1.22$\times 10^{-3}$ \\
W1.0020.000 & 26.74 & 36.3797 & -5.0612 & 0.5938 & 2.59$\times 10^{-18}$ \\
W1.0020.001 & 23.87 & 35.8023 & -4.6163 & 0.5708 & 3.07$\times 10^{-15}$ \\
W1.0020.002 & 22.02 & 35.1670 & -4.7508 & 0.5906 & 4.58$\times 10^{-13}$ \\
W1.0020.003 & 21.09 & 34.0741 & -5.1421 & 0.5875 & 7.46$\times 10^{-12}$ \\
W1.0020.004 & 20.86 & 36.9689 & -4.7235 & 0.5737 & 1.55$\times 10^{-11}$ \\
W1.0020.005 & 17.65 & 34.6769 & -5.0150 & 0.5777 & 7.19$\times 10^{-7}$ \\
W1.0020.006 & 17.10 & 37.4161 & -4.7254 & 0.5997 & 4.24$\times 10^{-6}$ \\
W1.0021.000 & 15.14 & 38.5026 & -4.6475 & 0.5939 & 7.79$\times 10^{-4}$ \\
W1.0022.000 & 23.68 & 33.2428 & -4.7257 & 0.6041 & 5.07$\times 10^{-15}$ \\
W1.0023.000 & 18.63 & 31.6109 & -4.4200 & 0.6042 & 2.59$\times 10^{-8}$ \\
W1.0024.000 & 16.57 & 32.0459 & -5.2097 & 0.6045 & 2.07$\times 10^{-5}$ \\
W1.0025.000 & 17.89 & 30.7988 & -5.1330 & 0.6050 & 3.21$\times 10^{-7}$ \\
W1.0026.000 & 22.00 & 34.5675 & -4.6252 & 0.6108 & 4.92$\times 10^{-13}$ \\
W1.0026.001 & 19.80 & 35.6312 & -4.8527 & 0.6087 & 4.82$\times 10^{-10}$ \\
W1.0026.002 & 19.25 & 33.6395 & -5.2028 & 0.6160 & 3.18$\times 10^{-9}$ \\
W1.0027.000 & 15.85 & 32.1356 & -4.7767 & 0.6162 & 1.48$\times 10^{-4}$ \\
W1.0028.000 & 14.97 & 30.6053 & -4.3257 & 0.6163 & 1.10$\times 10^{-3}$ \\
W1.0029.000 & 15.33 & 37.6681 & -4.3267 & 0.6180 & 5.08$\times 10^{-4}$ \\
W1.0030.000 & 14.94 & 36.8637 & -4.8765 & 0.6191 & 1.17$\times 10^{-3}$ \\
W1.0031.000 & 15.45 & 32.7694 & -5.2010 & 0.6192 & 3.93$\times 10^{-4}$ \\
W1.0032.000 & 14.80 & 38.1664 & -5.0000 & 0.6235 & 1.53$\times 10^{-3}$ \\
W1.0033.000 & 21.34 & 33.0056 & -4.4299 & 0.6249 & 3.42$\times 10^{-12}$ \\
W1.0033.001 & 16.89 & 33.3125 & -5.1943 & 0.6308 & 8.19$\times 10^{-6}$ \\
W1.0034.000 & 19.41 & 31.7580 & -4.8505 & 0.6293 & 1.79$\times 10^{-9}$ \\
W1.0034.001 & 18.08 & 30.8598 & -5.0710 & 0.6322 & 1.75$\times 10^{-7}$ \\
W1.0035.000 & 20.85 & 37.1440 & -5.1940 & 0.6315 & 1.58$\times 10^{-11}$ \\
W1.0036.000 & 16.04 & 38.1375 & -4.6544 & 0.6345 & 9.14$\times 10^{-5}$ \\
W1.0037.000 & 14.89 & 33.1509 & -4.3378 & 0.6367 & 1.29$\times 10^{-3}$ \\
W1.0038.000 & 14.96 & 34.5163 & -4.9228 & 0.6375 & 1.13$\times 10^{-3}$ \\
W1.0039.000 & 20.96 & 35.6851 & -5.1906 & 0.6377 & 1.11$\times 10^{-11}$ \\
W1.0040.000 & 17.30 & 35.0509 & -4.3392 & 0.6392 & 2.27$\times 10^{-6}$ \\
W1.0041.000 & 17.69 & 32.6090 & -4.8732 & 0.6443 & 6.40$\times 10^{-7}$ \\
W1.0042.000 & 15.49 & 31.5471 & -4.8008 & 0.6447 & 3.53$\times 10^{-4}$ \\
W1.0043.000 & 15.52 & 38.1678 & -5.2568 & 0.6485 & 3.30$\times 10^{-4}$ \\
W1.0044.000 & 20.93 & 33.7775 & -5.0386 & 0.6547 & 1.23$\times 10^{-11}$ \\
W1.0044.001 & 15.36 & 34.1561 & -4.3449 & 0.6491 & 4.80$\times 10^{-4}$ \\
W1.0045.000 & 20.95 & 31.3486 & -4.3482 & 0.6552 & 1.16$\times 10^{-11}$ \\
W1.0046.000 & 18.63 & 35.0847 & -5.1559 & 0.6583 & 2.67$\times 10^{-8}$ \\
W1.0047.000 & 15.79 & 30.5924 & -4.3988 & 0.6614 & 1.71$\times 10^{-4}$ \\
W1.0048.000 & 16.97 & 33.0296 & -4.9653 & 0.6624 & 6.24$\times 10^{-6}$ \\
W1.0049.000 & 18.05 & 30.7431 & -5.1771 & 0.6633 & 1.87$\times 10^{-7}$ \\
W1.0050.000 & 15.80 & 34.9392 & -4.4471 & 0.6642 & 1.67$\times 10^{-4}$ \\
W1.0051.000 & 15.41 & 33.7224 & -4.3568 & 0.6714 & 4.27$\times 10^{-4}$ \\
W1.0052.000 & 22.16 & 30.4256 & -4.8462 & 0.6751 & 3.13$\times 10^{-13}$ \\
W1.0053.000 & 24.82 & 33.2163 & -5.1013 & 0.6758 & 2.77$\times 10^{-16}$ \\
W1.0054.000 & 28.07 & 34.8620 & -4.6148 & 0.6786 & 1.02$\times 10^{-19}$ \\
W1.0054.001 & 25.36 & 37.1230 & -4.6345 & 0.6587 & 7.43$\times 10^{-17}$ \\
W1.0054.002 & 23.65 & 37.1926 & -4.5228 & 0.6802 & 5.49$\times 10^{-15}$ \\
W1.0054.003 & 22.40 & 36.1058 & -5.1326 & 0.6576 & 1.56$\times 10^{-13}$ \\
W1.0054.004 & 19.94 & 36.5885 & -4.6803 & 0.6457 & 3.04$\times 10^{-10}$ \\
W1.0054.005 & 19.26 & 37.8811 & -5.1737 & 0.6701 & 3.04$\times 10^{-9}$ \\
W1.0054.006 & 17.53 & 35.7799 & -4.3552 & 0.6683 & 1.06$\times 10^{-6}$ \\
W1.0054.007 & 17.41 & 37.5281 & -4.7285 & 0.6459 & 1.55$\times 10^{-6}$ \\
W1.0054.008 & 17.16 & 35.5437 & -5.2439 & 0.6699 & 3.46$\times 10^{-6}$ \\
W1.0054.009 & 16.19 & 33.9436 & -4.3880 & 0.6880 & 6.00$\times 10^{-5}$ \\
W1.0055.000 & 15.15 & 35.9250 & -4.5224 & 0.6789 & 7.54$\times 10^{-4}$ \\
W1.0056.000 & 15.63 & 33.2046 & -4.2938 & 0.6831 & 2.54$\times 10^{-4}$ \\
W1.0057.000 & 14.64 & 30.9997 & -4.4551 & 0.6841 & 2.05$\times 10^{-3}$ \\
W1.0058.000 & 18.28 & 33.6716 & -5.2324 & 0.6903 & 8.71$\times 10^{-8}$ \\
W1.0059.000 & 23.65 & 35.8813 & -5.1637 & 0.6912 & 5.41$\times 10^{-15}$ \\
W1.0060.000 & 18.80 & 30.3945 & -4.7993 & 0.6920 & 1.45$\times 10^{-8}$ \\
W1.0061.000 & 24.02 & 38.4943 & -4.9356 & 0.6930 & 2.07$\times 10^{-15}$ \\
W1.0061.001 & 22.18 & 37.7050 & -4.7316 & 0.7029 & 2.96$\times 10^{-13}$ \\
W1.0062.000 & 15.68 & 35.7151 & -4.3010 & 0.6958 & 2.27$\times 10^{-4}$ \\
W1.0063.000 & 23.37 & 31.3482 & -4.4148 & 0.6972 & 1.14$\times 10^{-14}$ \\
W1.0063.001 & 22.63 & 31.7741 & -4.8226 & 0.6819 & 8.39$\times 10^{-14}$ \\
W1.0064.000 & 19.35 & 32.3933 & -4.4832 & 0.6988 & 2.21$\times 10^{-9}$ \\
W1.0065.000 & 17.50 & 34.1766 & -4.9794 & 0.6996 & 1.19$\times 10^{-6}$ \\
W1.0066.000 & 16.64 & 34.9425 & -5.2269 & 0.7006 & 1.72$\times 10^{-5}$ \\
W1.0067.000 & 16.00 & 33.1911 & -4.3055 & 0.7039 & 1.01$\times 10^{-4}$ \\
W1.0068.000 & 16.32 & 30.3311 & -4.7095 & 0.7068 & 4.27$\times 10^{-5}$ \\
W1.0069.000 & 16.61 & 36.0683 & -4.3105 & 0.7131 & 1.86$\times 10^{-5}$ \\
W1.0070.000 & 16.88 & 32.4085 & -5.2198 & 0.7143 & 8.25$\times 10^{-6}$ \\
W1.0071.000 & 17.64 & 31.8043 & -4.3140 & 0.7197 & 7.45$\times 10^{-7}$ \\
W1.0072.000 & 16.58 & 38.3919 & -4.6885 & 0.7201 & 2.05$\times 10^{-5}$ \\
W1.0073.000 & 18.98 & 31.0840 & -5.2160 & 0.7217 & 7.96$\times 10^{-9}$ \\
W1.0074.000 & 14.79 & 35.3262 & -4.4035 & 0.7229 & 1.55$\times 10^{-3}$ \\
W1.0075.000 & 31.35 & 34.1867 & -5.0821 & 0.7275 & 3.30$\times 10^{-23}$ \\
W1.0075.001 & 25.66 & 35.4025 & -4.9985 & 0.7127 & 3.55$\times 10^{-17}$ \\
W1.0075.002 & 19.93 & 35.2743 & -4.9720 & 0.7321 & 3.10$\times 10^{-10}$ \\
W1.0075.003 & 19.82 & 33.1670 & -4.6661 & 0.7173 & 4.54$\times 10^{-10}$ \\
W1.0075.004 & 19.58 & 36.7327 & -5.0205 & 0.7137 & 1.03$\times 10^{-9}$ \\
W1.0075.005 & 16.93 & 35.0301 & -4.3059 & 0.7047 & 7.13$\times 10^{-6}$ \\
W1.0075.006 & 16.37 & 33.6268 & -4.3221 & 0.7355 & 3.68$\times 10^{-5}$ \\
W1.0075.007 & 15.71 & 34.4709 & -4.6475 & 0.7415 & 2.11$\times 10^{-4}$ \\
W1.0076.000 & 18.95 & 32.1151 & -5.2126 & 0.7287 & 8.81$\times 10^{-9}$ \\
W1.0077.000 & 14.64 & 33.2312 & -5.2766 & 0.7312 & 2.06$\times 10^{-3}$ \\
W1.0078.000 & 17.42 & 36.1006 & -4.3854 & 0.7317 & 1.51$\times 10^{-6}$ \\
W1.0079.000 & 15.32 & 30.7787 & -5.2105 & 0.7328 & 5.21$\times 10^{-4}$ \\
W1.0080.000 & 15.05 & 37.8517 & -4.4083 & 0.7345 & 9.31$\times 10^{-4}$ \\
W1.0081.000 & 16.69 & 32.8302 & -4.9060 & 0.7378 & 1.46$\times 10^{-5}$ \\
W1.0082.000 & 14.81 & 32.1626 & -4.3236 & 0.7385 & 1.49$\times 10^{-3}$ \\
W1.0083.000 & 23.33 & 31.6820 & -5.1411 & 0.7437 & 1.26$\times 10^{-14}$ \\
W1.0083.001 & 21.96 & 30.8265 & -4.8617 & 0.7499 & 5.55$\times 10^{-13}$ \\
W1.0083.002 & 19.84 & 30.6015 & -4.4093 & 0.7372 & 4.22$\times 10^{-10}$ \\
W1.0084.000 & 15.41 & 35.1357 & -5.2697 & 0.7437 & 4.24$\times 10^{-4}$ \\
W1.0085.000 & 28.39 & 37.4384 & -4.4778 & 0.7483 & 4.68$\times 10^{-20}$ \\
W1.0085.001 & 26.00 & 36.4852 & -5.0308 & 0.7545 & 1.57$\times 10^{-17}$ \\
W1.0085.002 & 22.69 & 37.5927 & -4.4624 & 0.7660 & 7.07$\times 10^{-14}$ \\
W1.0085.003 & 22.64 & 38.2412 & -5.0349 & 0.7401 & 8.02$\times 10^{-14}$ \\
W1.0085.004 & 17.54 & 35.9515 & -4.5619 & 0.7432 & 1.04$\times 10^{-6}$ \\
W1.0085.005 & 17.40 & 36.9162 & -4.9645 & 0.7682 & 1.64$\times 10^{-6}$ \\
W1.0085.006 & 17.21 & 37.1397 & -4.5162 & 0.7332 & 3.00$\times 10^{-6}$ \\
W1.0085.007 & 15.44 & 36.6871 & -4.8626 & 0.7409 & 4.01$\times 10^{-4}$ \\
W1.0086.000 & 16.13 & 34.3645 & -5.2029 & 0.7491 & 7.05$\times 10^{-5}$ \\
W1.0087.000 & 15.42 & 35.3661 & -4.3929 & 0.7495 & 4.14$\times 10^{-4}$ \\
W1.0088.000 & 17.55 & 33.4760 & -4.4589 & 0.7554 & 9.91$\times 10^{-7}$ \\
W1.0089.000 & 14.95 & 32.6976 & -4.2698 & 0.7582 & 1.13$\times 10^{-3}$ \\
W1.0090.000 & 15.63 & 38.3366 & -4.8609 & 0.7584 & 2.57$\times 10^{-4}$ \\
W1.0091.000 & 17.61 & 31.3893 & -4.5233 & 0.7587 & 8.23$\times 10^{-7}$ \\
W1.0092.000 & 15.38 & 30.4482 & -4.3977 & 0.7612 & 4.54$\times 10^{-4}$ \\
W1.0093.000 & 17.57 & 32.7896 & -5.1969 & 0.7623 & 9.26$\times 10^{-7}$ \\
W1.0094.000 & 22.81 & 34.2832 & -4.7555 & 0.7664 & 5.10$\times 10^{-14}$ \\
W1.0094.001 & 20.29 & 34.0232 & -4.9601 & 0.7910 & 3.77$\times 10^{-10}$ \\
W1.0094.002 & 19.43 & 34.6983 & -4.8793 & 0.7828 & 3.40$\times 10^{-9}$ \\
W1.0094.003 & 16.72 & 33.2479 & -5.1029 & 0.7912 & 3.63$\times 10^{-5}$ \\
W1.0095.000 & 16.14 & 38.4923 & -4.3371 & 0.7667 & 6.97$\times 10^{-5}$ \\
W1.0096.000 & 15.08 & 33.6569 & -4.2751 & 0.7682 & 8.66$\times 10^{-4}$ \\
W1.0097.000 & 21.91 & 32.6647 & -4.7142 & 0.7738 & 6.37$\times 10^{-13}$ \\
W1.0097.001 & 14.92 & 32.0546 & -5.0656 & 0.7791 & 1.43$\times 10^{-3}$ \\
W1.0098.000 & 17.50 & 35.2811 & -4.4237 & 0.7745 & 1.18$\times 10^{-6}$ \\
W1.0099.000 & 15.42 & 33.9120 & -5.2533 & 0.7755 & 4.42$\times 10^{-4}$ \\
W1.0100.000 & 17.80 & 36.0885 & -4.3413 & 0.7758 & 4.82$\times 10^{-7}$ \\
W1.0101.000 & 16.86 & 38.0965 & -5.2525 & 0.7770 & 1.05$\times 10^{-5}$ \\
W1.0102.000 & 15.40 & 34.3478 & -4.2803 & 0.7783 & 5.22$\times 10^{-4}$ \\
W1.0103.000 & 17.86 & 33.6059 & -4.3435 & 0.7808 & 5.76$\times 10^{-7}$ \\
W1.0104.000 & 25.03 & 31.2233 & -5.0025 & 0.7841 & 3.51$\times 10^{-16}$ \\
W1.0104.001 & 24.14 & 31.5409 & -4.3521 & 0.8007 & 1.16$\times 10^{-14}$ \\
W1.0104.002 & 23.36 & 30.4351 & -4.5344 & 0.8026 & 1.02$\times 10^{-13}$ \\
W1.0104.003 & 18.61 & 31.6950 & -5.2391 & 0.8052 & 3.04$\times 10^{-7}$ \\
W1.0104.004 & 15.87 & 30.4107 & -5.1863 & 0.7867 & 2.56$\times 10^{-4}$ \\
W1.0104.005 & 15.84 & 30.9781 & -4.2864 & 0.7903 & 3.24$\times 10^{-4}$ \\
W1.0105.000 & 16.53 & 31.7179 & -4.2834 & 0.7843 & 4.16$\times 10^{-5}$ \\
W1.0106.000 & 25.43 & 35.9663 & -4.8993 & 0.7880 & 1.81$\times 10^{-16}$ \\
W1.0107.000 & 16.67 & 32.6795 & -4.4098 & 0.7928 & 4.65$\times 10^{-5}$ \\
W1.0108.000 & 15.12 & 32.5899 & -4.9383 & 0.7992 & 2.00$\times 10^{-3}$ \\
W1.0109.000 & 17.35 & 37.0185 & -4.7562 & 0.8010 & 1.09$\times 10^{-5}$ \\
W1.0110.000 & 16.33 & 33.2202 & -4.3928 & 0.8014 & 1.70$\times 10^{-4}$ \\
W1.0111.000 & 15.41 & 32.6660 & -4.2948 & 0.8077 & 1.56$\times 10^{-3}$ \\
W1.0112.000 & 15.67 & 36.1465 & -4.2983 & 0.8151 & 1.24$\times 10^{-3}$ \\
W1.0113.000 & 22.68 & 33.7557 & -4.5374 & 0.8155 & 1.76$\times 10^{-12}$ \\
W1.0113.001 & 21.27 & 34.7733 & -4.8967 & 0.8086 & 6.79$\times 10^{-11}$ \\
W1.0113.002 & 17.41 & 33.9025 & -4.9534 & 0.8284 & 4.45$\times 10^{-5}$ \\
W1.0113.003 & 15.03 & 34.5736 & -4.4322 & 0.7987 & 2.29$\times 10^{-3}$ \\
W1.0114.000 & 29.16 & 36.8855 & -5.0927 & 0.8231 & 4.36$\times 10^{-19}$ \\
W1.0114.001 & 24.19 & 38.1743 & -5.0364 & 0.8114 & 2.23$\times 10^{-14}$ \\
W1.0114.002 & 22.44 & 36.7293 & -4.6597 & 0.8407 & 2.49$\times 10^{-11}$ \\
W1.0114.003 & 21.08 & 36.0996 & -5.0576 & 0.8068 & 1.07$\times 10^{-10}$ \\
W1.0114.004 & 20.54 & 38.3426 & -4.3467 & 0.7882 & 1.31$\times 10^{-10}$ \\
W1.0114.005 & 20.48 & 37.5508 & -5.1840 & 0.7923 & 2.29$\times 10^{-10}$ \\
W1.0114.006 & 20.37 & 37.5006 & -4.3555 & 0.8089 & 1.26$\times 10^{-9}$ \\
W1.0114.007 & 18.24 & 38.3383 & -5.1831 & 0.7946 & 4.66$\times 10^{-7}$ \\
W1.0114.008 & 17.30 & 37.4875 & -4.4894 & 0.7862 & 4.97$\times 10^{-6}$ \\
W1.0114.009 & 17.14 & 38.3006 & -4.3134 & 0.8045 & 2.47$\times 10^{-5}$ \\
W1.0114.010 & 16.09 & 37.7595 & -4.5128 & 0.7971 & 2.45$\times 10^{-4}$ \\
W1.0114.011 & 15.84 & 36.2263 & -5.2243 & 0.8389 & 2.05$\times 10^{-3}$ \\
W1.0114.012 & 15.66 & 37.3585 & -5.1971 & 0.8096 & 1.04$\times 10^{-3}$ \\
W1.0115.000 & 21.26 & 33.1529 & -4.3627 & 0.8268 & 3.00$\times 10^{-10}$ \\
W1.0115.001 & 17.97 & 32.9771 & -4.7964 & 0.8143 & 4.16$\times 10^{-6}$ \\
W1.0116.000 & 17.01 & 36.0490 & -4.3431 & 0.8269 & 1.14$\times 10^{-4}$ \\
W1.0117.000 & 16.87 & 31.4773 & -4.5400 & 0.8269 & 1.60$\times 10^{-4}$ \\
W1.0118.000 & 18.44 & 38.3492 & -4.3046 & 0.8289 & 2.59$\times 10^{-6}$ \\
W1.0119.000 & 15.88 & 32.2119 & -4.5012 & 0.8291 & 1.38$\times 10^{-3}$ \\
W1.0120.000 & 19.86 & 32.7171 & -4.9130 & 0.8371 & 5.72$\times 10^{-8}$ \\
W1.0121.000 & 15.78 & 32.8825 & -4.3109 & 0.8432 & 2.56$\times 10^{-3}$ \\
W1.0122.000 & 16.07 & 37.1467 & -5.2799 & 0.8450 & 1.68$\times 10^{-3}$ \\
W1.0123.000 & 17.40 & 34.9947 & -5.2204 & 0.8483 & 1.24$\times 10^{-4}$ \\
W1.0124.000 & 17.91 & 37.7340 & -4.3136 & 0.8496 & 3.80$\times 10^{-5}$ \\
W1.0125.000 & 18.71 & 35.6456 & -4.4690 & 0.8535 & 5.48$\times 10^{-6}$ \\
W1.0126.000 & 18.61 & 38.4312 & -4.7956 & 0.8544 & 7.68$\times 10^{-6}$ \\
W1.0127.000 & 16.37 & 30.9590 & -5.2754 & 0.8547 & 1.34$\times 10^{-3}$ \\
W1.0128.000 & 25.19 & 37.1348 & -4.7381 & 0.8578 & 5.29$\times 10^{-14}$ \\
W1.0128.001 & 20.15 & 37.8197 & -5.2236 & 0.8405 & 2.96$\times 10^{-8}$ \\
W1.0129.000 & 16.10 & 34.7413 & -4.9867 & 0.8588 & 2.42$\times 10^{-3}$ \\
W1.0130.000 & 29.32 & 33.1232 & -4.7574 & 0.8646 & 7.01$\times 10^{-18}$ \\
W1.0130.001 & 27.92 & 32.0782 & -4.8141 & 0.8742 & 2.94$\times 10^{-16}$ \\
W1.0130.002 & 27.44 & 31.6059 & -4.3724 & 0.8522 & 1.84$\times 10^{-16}$ \\
W1.0130.003 & 23.97 & 31.7576 & -5.1279 & 0.8360 & 2.44$\times 10^{-13}$ \\
W1.0130.004 & 23.55 & 34.1296 & -4.5088 & 0.8592 & 4.18$\times 10^{-12}$ \\
W1.0130.005 & 21.86 & 30.8975 & -4.5030 & 0.8361 & 9.83$\times 10^{-11}$ \\
W1.0130.006 & 21.18 & 32.1893 & -4.5349 & 0.8915 & 4.77$\times 10^{-8}$ \\
W1.0130.007 & 21.07 & 33.9890 & -5.0670 & 0.8446 & 2.18$\times 10^{-9}$ \\
W1.0130.008 & 18.45 & 31.2851 & -4.5487 & 0.8675 & 2.51$\times 10^{-5}$ \\
W1.0130.009 & 18.12 & 33.6180 & -4.4071 & 0.8411 & 1.38$\times 10^{-5}$ \\
W1.0130.010 & 17.82 & 31.5090 & -4.3290 & 0.8873 & 2.73$\times 10^{-4}$ \\
W1.0130.011 & 17.68 & 32.4745 & -4.4305 & 0.8532 & 8.21$\times 10^{-5}$ \\
W1.0130.012 & 16.74 & 30.7499 & -4.3137 & 0.8498 & 5.68$\times 10^{-4}$ \\
W1.0131.000 & 18.35 & 30.7312 & -5.1371 & 0.8667 & 3.07$\times 10^{-5}$ \\
W1.0132.000 & 22.70 & 38.3087 & -4.3780 & 0.8675 & 8.81$\times 10^{-11}$ \\
W1.0133.000 & 17.53 & 34.3601 & -5.2692 & 0.8683 & 2.28$\times 10^{-4}$ \\
W1.0134.000 & 17.91 & 37.6363 & -5.0981 & 0.8700 & 1.05$\times 10^{-4}$ \\
W1.0135.000 & 18.04 & 34.1159 & -4.3241 & 0.8750 & 9.93$\times 10^{-5}$ \\
W1.0136.000 & 16.58 & 33.6260 & -5.2662 & 0.8751 & 1.81$\times 10^{-3}$ \\
W1.0137.000 & 22.00 & 35.0765 & -4.3244 & 0.8757 & 1.30$\times 10^{-9}$ \\
W1.0138.000 & 24.35 & 30.5994 & -4.4381 & 0.8780 & 1.94$\times 10^{-12}$ \\
W1.0138.001 & 18.56 & 30.8528 & -5.2076 & 0.8806 & 3.73$\times 10^{-5}$ \\
W1.0139.000 & 24.03 & 36.0749 & -5.1858 & 0.8892 & 1.01$\times 10^{-11}$ \\
W1.0139.001 & 19.89 & 35.3159 & -5.0195 & 0.8833 & 1.24$\times 10^{-6}$ \\
W1.0139.002 & 17.54 & 35.7915 & -4.5505 & 0.8764 & 3.17$\times 10^{-4}$ \\
W1.0140.000 & 22.17 & 38.2791 & -5.0360 & 0.8931 & 2.81$\times 10^{-9}$ \\
W1.0141.000 & 19.59 & 34.3076 & -4.9801 & 0.8944 & 5.51$\times 10^{-6}$ \\
W1.0142.000 & 18.53 & 31.3280 & -4.9983 & 0.8960 & 8.42$\times 10^{-5}$ \\
W1.0143.000 & 19.17 & 33.2938 & -4.3337 & 0.8994 & 2.17$\times 10^{-5}$ \\
W1.0144.000 & 17.57 & 37.2608 & -5.2559 & 0.8994 & 7.12$\times 10^{-4}$ \\
W1.0145.000 & 17.58 & 34.8983 & -4.2787 & 0.9004 & 7.23$\times 10^{-4}$ \\
W4.0001.000 & 20.13 & 334.5530 & 1.1434 & 0.5498 & 1.64$\times 10^{-10}$ \\
W4.0001.001 & 14.75 & 334.3317 & 1.1786 & 0.5608 & 1.70$\times 10^{-3}$ \\
W4.0002.000 & 17.21 & 332.3445 & 2.2182 & 0.5524 & 3.02$\times 10^{-6}$ \\
W4.0003.000 & 14.63 & 333.6647 & 1.4755 & 0.5527 & 2.11$\times 10^{-3}$ \\
W4.0004.000 & 15.04 & 330.3109 & 1.4206 & 0.5531 & 9.44$\times 10^{-4}$ \\
W4.0005.000 & 17.00 & 334.0628 & 2.2115 & 0.5596 & 5.77$\times 10^{-6}$ \\
W4.0006.000 & 22.81 & 330.3673 & 1.7740 & 0.5686 & 5.07$\times 10^{-14}$ \\
W4.0007.000 & 24.24 & 332.1927 & 1.1033 & 0.5688 & 1.20$\times 10^{-15}$ \\
W4.0007.001 & 24.19 & 333.3963 & 2.1200 & 0.5725 & 1.35$\times 10^{-15}$ \\
W4.0007.002 & 16.89 & 332.7841 & 1.4768 & 0.5587 & 8.19$\times 10^{-6}$ \\
W4.0008.000 & 23.06 & 334.8589 & 1.4268 & 0.5741 & 2.57$\times 10^{-14}$ \\
W4.0009.000 & 15.98 & 330.2894 & 1.5094 & 0.5915 & 1.05$\times 10^{-4}$ \\
W4.0010.000 & 16.37 & 335.0833 & 1.5614 & 0.5936 & 3.75$\times 10^{-5}$ \\
W4.0011.000 & 25.24 & 332.6775 & 1.7673 & 0.5985 & 9.99$\times 10^{-17}$ \\
W4.0011.001 & 20.42 & 332.1264 & 1.7703 & 0.5846 & 6.35$\times 10^{-11}$ \\
W4.0011.002 & 18.30 & 332.8595 & 1.0912 & 0.5875 & 8.07$\times 10^{-8}$ \\
W4.0011.003 & 16.40 & 333.4901 & 2.0447 & 0.6070 & 3.35$\times 10^{-5}$ \\
W4.0012.000 & 18.97 & 334.3651 & 1.0508 & 0.6019 & 8.24$\times 10^{-9}$ \\
W4.0013.000 & 16.79 & 334.5128 & 1.9445 & 0.6042 & 1.10$\times 10^{-5}$ \\
W4.0014.000 & 23.57 & 331.8038 & 2.0131 & 0.6182 & 6.68$\times 10^{-15}$ \\
W4.0014.001 & 21.79 & 331.2497 & 1.0523 & 0.6037 & 8.99$\times 10^{-13}$ \\
W4.0014.002 & 21.70 & 330.7721 & 1.4377 & 0.6157 & 1.20$\times 10^{-12}$ \\
W4.0014.003 & 19.29 & 331.0512 & 2.1558 & 0.6281 & 2.69$\times 10^{-9}$ \\
W4.0014.004 & 16.14 & 330.2542 & 1.7864 & 0.6258 & 6.90$\times 10^{-5}$ \\
W4.0015.000 & 16.70 & 332.0323 & 1.0663 & 0.6220 & 1.44$\times 10^{-5}$ \\
W4.0016.000 & 16.32 & 334.8098 & 1.4399 & 0.6247 & 4.24$\times 10^{-5}$ \\
W4.0017.000 & 19.76 & 333.4463 & 2.1579 & 0.6252 & 5.46$\times 10^{-10}$ \\
W4.0017.001 & 18.31 & 333.5599 & 1.0591 & 0.6125 & 7.86$\times 10^{-8}$ \\
W4.0017.002 & 16.96 & 334.4585 & 1.7108 & 0.6370 & 6.46$\times 10^{-6}$ \\
W4.0017.003 & 15.24 & 333.1045 & 1.7385 & 0.6162 & 6.23$\times 10^{-4}$ \\
W4.0018.000 & 17.54 & 333.9272 & 1.0751 & 0.6340 & 1.03$\times 10^{-6}$ \\
W4.0019.000 & 23.14 & 332.3126 & 1.0754 & 0.6345 & 2.09$\times 10^{-14}$ \\
W4.0020.000 & 15.07 & 331.4354 & 1.5158 & 0.6363 & 8.98$\times 10^{-4}$ \\
W4.0021.000 & 15.06 & 330.8152 & 1.1534 & 0.6420 & 9.09$\times 10^{-4}$ \\
W4.0022.000 & 26.37 & 332.8245 & 2.1433 & 0.6458 & 6.27$\times 10^{-18}$ \\
W4.0022.001 & 16.18 & 333.7578 & 1.9007 & 0.6505 & 6.17$\times 10^{-5}$ \\
W4.0023.000 & 15.61 & 330.6889 & 1.9732 & 0.6492 & 2.65$\times 10^{-4}$ \\
W4.0024.000 & 17.03 & 330.5763 & 1.0158 & 0.6520 & 5.25$\times 10^{-6}$ \\
W4.0025.000 & 17.49 & 332.1786 & 1.0162 & 0.6524 & 1.24$\times 10^{-6}$ \\
W4.0026.000 & 19.83 & 334.5600 & 1.5658 & 0.6557 & 4.41$\times 10^{-10}$ \\
W4.0027.000 & 16.03 & 331.0046 & 1.2830 & 0.6618 & 9.30$\times 10^{-5}$ \\
W4.0028.000 & 19.46 & 331.9753 & 1.4018 & 0.6647 & 1.52$\times 10^{-9}$ \\
W4.0029.000 & 20.90 & 333.7386 & 1.0964 & 0.6652 & 1.35$\times 10^{-11}$ \\
W4.0030.000 & 14.87 & 331.2633 & 2.1945 & 0.6743 & 1.34$\times 10^{-3}$ \\
W4.0031.000 & 15.55 & 332.1940 & 1.1959 & 0.6756 & 3.05$\times 10^{-4}$ \\
W4.0032.000 & 16.11 & 330.2687 & 2.1932 & 0.6761 & 7.48$\times 10^{-5}$ \\
W4.0033.000 & 14.96 & 334.7357 & 1.4289 & 0.6801 & 1.11$\times 10^{-3}$ \\
W4.0034.000 & 15.76 & 334.0602 & 1.0622 & 0.6838 & 1.88$\times 10^{-4}$ \\
W4.0035.000 & 16.18 & 331.9482 & 1.0405 & 0.6856 & 6.27$\times 10^{-5}$ \\
W4.0036.000 & 16.42 & 330.5077 & 2.1855 & 0.6872 & 3.18$\times 10^{-5}$ \\
W4.0037.000 & 14.72 & 331.1795 & 0.9762 & 0.6915 & 1.80$\times 10^{-3}$ \\
W4.0038.000 & 20.27 & 333.6565 & 2.1800 & 0.6953 & 1.02$\times 10^{-10}$ \\
W4.0039.000 & 25.80 & 332.6369 & 1.9079 & 0.6958 & 2.52$\times 10^{-17}$ \\
W4.0039.001 & 25.13 & 331.4491 & 1.9103 & 0.6891 & 1.30$\times 10^{-16}$ \\
W4.0039.002 & 23.64 & 330.6290 & 1.3829 & 0.6807 & 5.56$\times 10^{-15}$ \\
W4.0039.003 & 20.50 & 330.3248 & 1.4549 & 0.6952 & 4.85$\times 10^{-11}$ \\
W4.0039.004 & 18.15 & 333.1144 & 1.0673 & 0.6915 & 1.37$\times 10^{-7}$ \\
W4.0039.005 & 16.80 & 331.4341 & 1.7696 & 0.7091 & 1.05$\times 10^{-5}$ \\
W4.0039.006 & 15.91 & 330.7968 & 1.8605 & 0.7033 & 1.28$\times 10^{-4}$ \\
W4.0040.000 & 15.98 & 334.7129 & 1.2977 & 0.6990 & 1.05$\times 10^{-4}$ \\
W4.0041.000 & 18.27 & 332.0578 & 1.1450 & 0.7083 & 8.89$\times 10^{-8}$ \\
W4.0042.000 & 18.82 & 334.3541 & 1.6359 & 0.7096 & 1.37$\times 10^{-8}$ \\
W4.0043.000 & 15.22 & 332.9044 & 1.7244 & 0.7146 & 6.49$\times 10^{-4}$ \\
W4.0044.000 & 17.50 & 333.4557 & 2.2317 & 0.7178 & 1.17$\times 10^{-6}$ \\
W4.0045.000 & 17.93 & 330.8160 & 1.8333 & 0.7223 & 2.81$\times 10^{-7}$ \\
W4.0046.000 & 16.23 & 335.1173 & 1.5698 & 0.7238 & 5.52$\times 10^{-5}$ \\
W4.0047.000 & 14.87 & 333.6026 & 1.3291 & 0.7252 & 1.34$\times 10^{-3}$ \\
W4.0048.000 & 20.95 & 332.9448 & 1.8532 & 0.7299 & 1.15$\times 10^{-11}$ \\
W4.0048.001 & 19.40 & 332.1431 & 1.7867 & 0.7360 & 1.89$\times 10^{-9}$ \\
W4.0048.002 & 15.56 & 332.5097 & 1.1790 & 0.7316 & 2.99$\times 10^{-4}$ \\
W4.0049.000 & 24.45 & 334.4038 & 1.1564 & 0.7300 & 7.11$\times 10^{-16}$ \\
W4.0049.001 & 20.25 & 334.8041 & 1.6351 & 0.7452 & 1.08$\times 10^{-10}$ \\
W4.0049.002 & 14.86 & 335.1018 & 1.0094 & 0.7379 & 1.37$\times 10^{-3}$ \\
W4.0050.000 & 24.69 & 330.7971 & 1.0717 & 0.7338 & 3.91$\times 10^{-16}$ \\
W4.0050.001 & 20.59 & 330.3501 & 1.5709 & 0.7453 & 3.67$\times 10^{-11}$ \\
W4.0051.000 & 17.35 & 333.9839 & 2.2204 & 0.7342 & 1.91$\times 10^{-6}$ \\
W4.0052.000 & 15.10 & 330.7381 & 2.1972 & 0.7366 & 8.36$\times 10^{-4}$ \\
W4.0053.000 & 16.96 & 331.6690 & 1.0092 & 0.7376 & 6.58$\times 10^{-6}$ \\
W4.0054.000 & 18.94 & 331.4193 & 1.7209 & 0.7441 & 9.12$\times 10^{-9}$ \\
W4.0055.000 & 16.15 & 333.9822 & 2.0179 & 0.7510 & 6.77$\times 10^{-5}$ \\
W4.0056.000 & 14.77 & 332.1350 & 0.9587 & 0.7575 & 1.62$\times 10^{-3}$ \\
W4.0057.000 & 16.54 & 330.8373 & 1.6977 & 0.7637 & 2.27$\times 10^{-5}$ \\
W4.0058.000 & 22.70 & 331.8457 & 1.6767 & 0.7643 & 6.75$\times 10^{-14}$ \\
W4.0058.001 & 20.46 & 331.3086 & 1.0211 & 0.7559 & 5.50$\times 10^{-11}$ \\
W4.0058.002 & 18.09 & 331.7175 & 1.4491 & 0.7822 & 3.02$\times 10^{-7}$ \\
W4.0058.003 & 14.80 & 331.7490 & 1.0106 & 0.7725 & 1.53$\times 10^{-3}$ \\
W4.0059.000 & 15.32 & 330.3174 & 1.2577 & 0.7658 & 5.20$\times 10^{-4}$ \\
W4.0060.000 & 28.12 & 332.9143 & 1.8432 & 0.7698 & 9.12$\times 10^{-20}$ \\
W4.0060.001 & 19.42 & 333.9559 & 1.9471 & 0.7708 & 1.74$\times 10^{-9}$ \\
W4.0060.002 & 18.66 & 334.6720 & 1.6966 & 0.7773 & 2.97$\times 10^{-8}$ \\
W4.0060.003 & 16.83 & 333.1196 & 0.9725 & 0.7775 & 1.20$\times 10^{-5}$ \\
W4.0060.004 & 15.21 & 333.4866 & 1.6348 & 0.7584 & 6.59$\times 10^{-4}$ \\
W4.0060.005 & 14.99 & 333.5513 & 1.7581 & 0.7807 & 1.33$\times 10^{-3}$ \\
W4.0061.000 & 14.79 & 335.1869 & 1.4679 & 0.7704 & 1.55$\times 10^{-3}$ \\
W4.0062.000 & 16.82 & 333.7450 & 1.1358 & 0.7731 & 9.99$\times 10^{-6}$ \\
W4.0063.000 & 16.72 & 331.7766 & 2.2539 & 0.7794 & 1.84$\times 10^{-5}$ \\
W4.0064.000 & 15.74 & 332.4376 & 1.3249 & 0.7800 & 2.52$\times 10^{-4}$ \\
W4.0065.000 & 18.35 & 330.8333 & 1.0377 & 0.7829 & 1.35$\times 10^{-7}$ \\
W4.0066.000 & 16.30 & 335.0924 & 0.9762 & 0.7832 & 7.38$\times 10^{-5}$ \\
W4.0067.000 & 17.33 & 332.4386 & 2.1895 & 0.7835 & 3.79$\times 10^{-6}$ \\
W4.0068.000 & 16.05 & 332.6652 & 1.2265 & 0.7919 & 2.14$\times 10^{-4}$ \\
W4.0069.000 & 22.75 & 330.2619 & 1.9803 & 0.7932 & 2.57$\times 10^{-13}$ \\
W4.0069.001 & 18.11 & 331.1608 & 2.1627 & 0.7948 & 7.16$\times 10^{-7}$ \\
W4.0069.002 & 14.89 & 330.8970 & 2.0454 & 0.7840 & 1.83$\times 10^{-3}$ \\
W4.0070.000 & 15.73 & 332.0001 & 0.9886 & 0.8026 & 6.99$\times 10^{-4}$ \\
W4.0071.000 & 15.78 & 332.0024 & 2.2170 & 0.8058 & 7.09$\times 10^{-4}$ \\
W4.0072.000 & 18.65 & 334.2292 & 2.2350 & 0.8093 & 3.61$\times 10^{-7}$ \\
W4.0073.000 & 16.08 & 332.8064 & 2.2350 & 0.8093 & 4.36$\times 10^{-4}$ \\
W4.0074.000 & 17.76 & 330.2413 & 2.1748 & 0.8095 & 5.73$\times 10^{-6}$ \\
W4.0075.000 & 20.00 & 334.4277 & 1.4738 & 0.8104 & 4.65$\times 10^{-9}$ \\
W4.0076.000 & 22.07 & 333.4841 & 1.5740 & 0.8139 & 9.32$\times 10^{-12}$ \\
W4.0076.001 & 20.60 & 333.8455 & 1.4307 & 0.7938 & 1.72$\times 10^{-10}$ \\
W4.0076.002 & 15.66 & 333.2311 & 1.4321 & 0.8014 & 7.72$\times 10^{-4}$ \\
W4.0077.000 & 19.37 & 332.5471 & 0.9970 & 0.8164 & 5.89$\times 10^{-8}$ \\
W4.0078.000 & 18.00 & 334.0211 & 2.2222 & 0.8310 & 1.07$\times 10^{-5}$ \\
W4.0079.000 & 20.45 & 333.3131 & 1.8684 & 0.8337 & 6.64$\times 10^{-9}$ \\
W4.0080.000 & 22.29 & 334.1325 & 1.1647 & 0.8358 & 2.65$\times 10^{-11}$ \\
W4.0080.001 & 15.59 & 334.0692 & 1.5547 & 0.8242 & 1.95$\times 10^{-3}$ \\
W4.0081.000 & 16.30 & 331.7152 & 1.0114 & 0.8411 & 9.55$\times 10^{-4}$ \\
W4.0082.000 & 17.67 & 334.2781 & 2.0795 & 0.8431 & 4.98$\times 10^{-5}$ \\
W4.0083.000 & 21.06 & 332.3785 & 1.3052 & 0.8489 & 3.14$\times 10^{-9}$ \\
W4.0084.000 & 18.14 & 335.1766 & 1.3439 & 0.8491 & 2.02$\times 10^{-5}$ \\
W4.0085.000 & 17.57 & 333.8008 & 1.8251 & 0.8545 & 1.14$\times 10^{-4}$ \\
W4.0086.000 & 15.99 & 330.3849 & 2.1127 & 0.8552 & 2.60$\times 10^{-3}$ \\
W4.0087.000 & 16.40 & 330.6731 & 0.9619 & 0.8555 & 1.32$\times 10^{-3}$ \\
W4.0088.000 & 26.97 & 331.3087 & 1.5182 & 0.8572 & 7.68$\times 10^{-16}$ \\
W4.0088.001 & 25.71 & 331.4488 & 1.5353 & 0.8296 & 2.02$\times 10^{-15}$ \\
W4.0088.002 & 24.67 & 332.3552 & 2.1603 & 0.8370 & 4.36$\times 10^{-14}$ \\
W4.0088.003 & 24.52 & 331.2612 & 1.7343 & 0.8078 & 7.37$\times 10^{-15}$ \\
W4.0088.004 & 23.33 & 330.5556 & 1.6333 & 0.8446 & 2.62$\times 10^{-12}$ \\
W4.0088.005 & 20.28 & 330.4122 & 2.1824 & 0.8320 & 1.02$\times 10^{-8}$ \\
W4.0088.006 & 19.32 & 331.6915 & 0.9971 & 0.8165 & 7.07$\times 10^{-8}$ \\
W4.0088.007 & 19.23 & 330.7883 & 2.0508 & 0.8663 & 2.81$\times 10^{-6}$ \\
W4.0088.008 & 18.84 & 331.1669 & 2.2180 & 0.8384 & 1.45$\times 10^{-6}$ \\
W4.0088.009 & 18.24 & 331.9934 & 2.0693 & 0.8215 & 2.93$\times 10^{-6}$ \\
W4.0088.010 & 17.96 & 331.1412 & 2.2081 & 0.8211 & 6.46$\times 10^{-6}$ \\
W4.0088.011 & 17.55 & 330.4278 & 1.3979 & 0.8301 & 3.42$\times 10^{-5}$ \\
W4.0088.012 & 16.68 & 330.4552 & 1.4933 & 0.8065 & 9.23$\times 10^{-5}$ \\
W4.0088.013 & 16.24 & 331.6637 & 2.1161 & 0.8479 & 1.37$\times 10^{-3}$ \\
W4.0089.000 & 18.18 & 332.2859 & 2.2069 & 0.8584 & 3.02$\times 10^{-5}$ \\
W4.0090.000 & 21.31 & 333.3950 & 1.0213 & 0.8588 & 3.00$\times 10^{-9}$ \\
W4.0091.000 & 18.00 & 334.1974 & 1.0216 & 0.8595 & 5.12$\times 10^{-5}$ \\
W4.0092.000 & 17.07 & 333.7134 & 2.2597 & 0.8662 & 5.45$\times 10^{-4}$ \\
W4.0093.000 & 23.48 & 335.1837 & 1.2738 & 0.8715 & 1.26$\times 10^{-11}$ \\
W4.0093.001 & 22.72 & 334.5934 & 1.7474 & 0.8626 & 5.69$\times 10^{-11}$ \\
W4.0094.000 & 17.76 & 331.8023 & 1.5198 & 0.8750 & 1.85$\times 10^{-4}$ \\
W4.0095.000 & 17.55 & 334.4866 & 2.1385 & 0.8818 & 3.83$\times 10^{-4}$ \\
W4.0096.000 & 21.56 & 335.0903 & 1.0370 & 0.8890 & 1.31$\times 10^{-8}$ \\
W4.0097.000 & 27.92 & 333.0036 & 2.0034 & 0.8935 & 1.10$\times 10^{-15}$ \\
W4.0097.001 & 24.39 & 332.7635 & 1.3490 & 0.8697 & 9.64$\times 10^{-13}$ \\
W4.0097.002 & 21.47 & 332.3672 & 2.1046 & 0.8734 & 5.49$\times 10^{-9}$ \\
W4.0097.003 & 19.29 & 333.5518 & 1.9707 & 0.8791 & 5.12$\times 10^{-6}$ \\
W4.0097.004 & 18.78 & 332.2810 & 1.5214 & 0.8947 & 4.42$\times 10^{-5}$ \\
W4.0097.005 & 17.74 & 332.1877 & 2.2446 & 0.8932 & 4.07$\times 10^{-4}$ \\
W4.0098.000 & 29.38 & 330.5937 & 1.7067 & 0.8939 & 4.90$\times 10^{-17}$ \\
W4.0098.001 & 22.49 & 330.3510 & 1.4444 & 0.8751 & 2.90$\times 10^{-10}$ \\
W4.0098.002 & 19.92 & 331.2488 & 1.2776 & 0.8842 & 1.20$\times 10^{-6}$ \\
W4.0099.000 & 17.20 & 333.6324 & 0.9848 & 0.8955 & 1.20$\times 10^{-3}$ \\
W4.0100.000 & 18.14 & 334.1837 & 2.0386 & 0.8988 & 2.28$\times 10^{-4}$ \\
\hline
\end{longtable}
\end{longtab}
\twocolumn

\end{document}